\documentclass{raa_twocolumn}     

\usepackage{graphicx,times}
\usepackage{natbib}
\usepackage{amssymb,amsmath}
\bibpunct{(}{)}{;}{a}{}{,}

\usepackage[pagebackref=true]{hyperref}

\begin{document}

\title{Cluster analysis of the Roma-BZCAT blazars}

\volnopage{ {\bf 20XX} Vol.\ {\bf X} No. {\bf XX}, 000--000}
\setcounter{page}{1}

\author{
Dmitry~O.~Kudryavtsev,\!\inst{1}
Yulia~V.~Sotnikova,\!\inst{1}
Vladislav~A.~Stolyarov,\!\inst{1,2}
Timur~V.~Mufakharov,\!\inst{1,3}
Valery~V.~Vlasyuk,\!\inst{1}
Margarita~L.~Khabibullina,\!\inst{1}
Alexander~G.~Mikhailov,\!\inst{1}
Yulia~V.~Cherepkova\inst{1}
}


\institute{
Special Astrophysical Observatory of the Russian Academy of Sciences, Nizhny Arkhyz 369167, Russia; {\it dkudr@sao.ru}\\
\and
Astrophysics Group, Cavendish Laboratory, University of Cambridge, Cambridge CB3~0HE, UK\\
\and 
Kazan Federal University, 18 Kremlyovskaya St, Kazan 420008, Russia\\
\vs \no
{\small Received 20XX Month Day; accepted 20XX Month Day}
}

\abstract{Based on the collected multiwavelength data, namely in the radio (NVSS, FIRST, RATAN-600), IR~(WISE), optical (Pan-STARRS), UV (GALEX), and X-ray (ROSAT, Swift-XRT) ranges, we have performed a cluster analysis for the blazars of the Roma-BZCAT catalog. Using two machine learning methods, namely a combination of PCA with k-means clustering and Kohonen's self-organizing maps, we have constructed an independent classification of the blazars (five classes) and compared the classes with the known Roma-BZCAT classification (FSRQs, BL\,Lacs, galaxy-dominated BL\,Lacs, and blazars of an uncertain type) as well as with the high synchrotron peaked blazars (HSP) from the 3HSP catalog and blazars from the TeVCat catalog. The obtained groups demonstrate concordance with the BL\,Lac/FSRQ classification along with a continuous character of the change in the properties. The group of HSP blazars stands out against the overall distribution. We examine the characteristics of the five groups and demonstrate distinctions in their spectral energy distribution shapes. The effectiveness of the clustering technique for objective analysis of multiparametric arrays of experimental data is demonstrated.
}

\authorrunning{D.~O.~Kudryavtsev et al.} 
\titlerunning{Cluster analysis of the Roma-BZCAT blazars} 
\maketitle

\section{Introduction}

Blazars are a rare type of active galactic nuclei (AGNs) with a jet of relativistic plasma pointing toward the Earth at relatively small angle (e.g., \citealt{1995PASP..107..803U, 2019ARA&A..57..467B}). 
Blazars are also among the brightest AGNs, Doppler-beaming effect \citep{1987MNRAS.224..257M, 1993ApJ...407...65G, 2017ApJ...835L..38F} makes their jet emission even more boosted and visible up to $z\sim6$ \citep{2020A&A...635L...7B}.
They are characterized by complex properties such as extreme variability at all wavelengths, high luminosity, high degree of polarization and brightness temperatures exceeding the Compton limit \citep{1999APh....11..159U}.

Blazars are the dominant sources in the extragalactic gamma-ray sky. Because of the relativistic amplification of their emission, sources even at high redshifts are observed.
The investigation of the multiwavelength properties of high redshift blazars is especially important as they are the most powerful non-explosive astrophysical sources and their study can be crucial for understanding the jet formation and propagation around supermassive black holes. Recently found connection between blazars and IceCube sources of high-energy neutrinos \citep{2020ApJ...894..101P} also adds to the topicality of their investigation.

The typical spectral energy distribution (SED) of a blazar is dominated by the non-thermal radiation from the jet and consists of the synchrotron (peaking between the far-infrared and the soft X-ray bands) and inverse-Compton (peaking in the hard X-ray to gamma-ray bands) humps \citep{2010ApJ...716...30A}. Besides that, the SED of a blazar can also feature the thermal radiation from the host galaxy (infrared hump or stellar emission) and the emission from the accretion disk around the central black hole (``blue hump'') and from the broad line region \citep{2012A&A...541A.160G}. Blazars exhibit large and rapid variations on a variety of time scales from years to intervals even shorter than an hour (e.g., \citealt{2017A&ARv..25....2P} and references therein).

Blazars are subclassified as flat-spectrum radio quasars (FSRQs) and BL\,Lacertae-type objects (BL\,Lacs) based on their optical spectra: FSRQs show broad emission lines, while BL\,Lacs display either very weak emission lines or even completely featureless (e.g., \citealt{1995PASP..107..803U,2014A&ARv..22...73F}). 
Another classification was proposed based on the luminosity of the broad-line region (BLR) in Eddington luminosity \citep{2011MNRAS.414.2674G}: sources with  $L_{\rm BLR} /L_{\rm Edd}$ higher or lower than $5\times 10^{-4}$ were classified as FSRQ or BL\,Lac, respectively, according to a transition of the accretion regime from radiatively efficient to inefficient one between the two classes.

Based on the peak frequency ($\nu_{\rm peak}$) of the synchrotron energy hump, blazars are usually subclassified as low (LSP, $\nu_{\rm peak} < 10^{14}$~Hz), intermediate (ISP, $10^{14}$~Hz $ <\nu_{\rm peak}< 10^{15} $~Hz), and high-synchrotron-peaked (HSP, $\nu_{\rm peak} > 10^{15}$ Hz) blazars \citep{2010ApJ...716...30A, 2016ApJS..226...20F}. Most HSP and ISP blazars have been classified as BL\,Lacs, while the LSP class contains both FSRQs and LSP BL\,Lacs \citep{2019Galax...7...20B, 2022Galax..10...35P}.

Inspired by the observed data, alternative physical categorizations for blazars are proposed: for instance, based on the sources with intrinsically weak or strong O\,II and O\,III emission lines \citep{2004MNRAS.351...83L}; based on the different accretion rates (the luminosity of the broad line region relative to the Eddington luminosity) of the two subclasses of blazars 
(e.g., \citealt{2011MNRAS.414.2674G, 2012MNRAS.421.1764S}); based on the ionizing radiation emitted from the accretion disc \citep{2012MNRAS.420.2899G, 2013MNRAS.431.1914G, 2015MNRAS.450.2404G}; 
based on the kinematic features of radio jets (e.g., \citealt{{2016A&A...592A..22H}}); etc.

The above mentioned numerous approaches to blazar classification tend to use a single categorical (presence/absence of emission lines) or a single numerical parameter (HSPs, $L_{\rm BLR} /L_{\rm Edd}$), in the latter case also categorized by setting a threshold defined by the researchers. At the same time blazars, like any objects, have numerous measurable characteristics that define their properties, and contemporary computing power and machine learning methods allows us to investigate a large number of characteristics in all their complexity.

In this paper we perform multiparametric cluster analysis for the Roma-BZCAT catalog \citep{2015ApSS.357...75M}, a sample of blazars with the most complete set of characteristics observed in different ranges of the electromagnetic spectrum. The aim is to divide the blazars into groups with similar properties to further analyze the differences between the groups to check the performance of the machine learning (clustering) methodology and compare it with the generally accepted classification approaches.

\section{General conception of cluster~analysis}

Cluster analysis, or clustering, is a classical problem of unsupervised machine learning (ML), i.e. learning with unlabelled data, when the model is not given in advance any target variable, in this case the classes of considered objects. The aim of the clustering model is to combine similar objects in groups (clusters) based on the similarity of their characteristics, or features. The principal idea is that when these characteristics are expressed numerically, the objects with similar properties are located closer to each other in the feature space than those with greater differences. In the simplest case of two--three features and clearly separated clusters this problem can be solved visually by constructing usual 2D or 3D scatter diagrams. In the general case of an arbitrary number of characteristics, the clustering must be performed in an $n$-dimensional feature space. ML algorithms are capable of solving such problems successfully even for complex distributions.

Notice that clustering in terms of machine learning should be distinguished from classification, which is a separate problem of supervised ML, when the model is trained to guess the classes known a priori. The main difference between the unsupervised problem of clustering and the supervised or semi-supervised problem of classification is the approach itself: while in the latter case we exploit a known classification developed by some other methods and assign the known classes to new objects, in clustering we develop a new classification based solely on the data collected for the objects.
This allows one to describe a sample based on experimental data, avoiding as much as possible the subjective approach to the division of objects into different types.

The mathematical formulation of the clustering is as follows.
Let $\{X\} \in \mathbb{R}^{N\times M}$ is a set of dimension $N\times M$, where $N$ is the number of objects ${\bf x}_n$ and $M$ is the number of their features ${\bf x}_m$. The set $\{X\}$ can be represented as a matrix ${\bf X} = (x_{nm})$ with \mbox{$n=1, 2, ..., N \equiv [1, N]$}, \mbox{$m = [1, M]$}. Let $\{Y\} \in \mathbb {Z}^K$ is a set of cardinality $K$, where $K$ is the number of clusters, \mbox{$\{Y\} = [1, K]$}. The solution of the clustering problem is finding an algorithmic function \mbox{$a$: $\{X\} \rightarrow \{Y\}$} that assign a singular label $y_k$, $k = [1, K]$ to each object ${\bf x}_n$, $n=[1, N]$
in such a way that the objects with similar properties (ideally forming separated groups in the feature space ${\bf x}_m$) correspond to the same label (cluster). Each object ${\bf x}_n$ can be represented in the feature space as a vector of dimension $M$: ${\bf x}_n = (x_{n1}, ..., x_{nM})$. The measure of object similarity is a metric of distance between the vectors, in our case this is the Euclidean distance: 
\begin{equation}
d=|{\bf x}_{i} - {\bf x}_{j}|,
\end{equation}
where ${\bf x}_i$ and ${\bf x}_j$ are any two vectors ${\bf x}_n$ (sample objects). In order for the features to have equal priorities in the clustering process, they should be normalized beforehand to the same scale. 

We should note that cluster analysis is a heuristic and its exact results are always model dependent both on the choice of the features selected for the modeling and on the clustering algorithm \mbox{$a$: $X \rightarrow Y$}.
An additional degree of freedom is the number of clusters $K$, which in most algorithms is defined a priori and as well evaluated heuristically based on the data. In this sense the obtained structure of the clusters should not be considered as an established natural phenomenon, especially when the clusters are not well separated; the cluster analysis, to a greater degree, is an instrument to search for patterns in the sample rather than investigation of individual objects.

Generally, the problem solution can be divided into several stages:
\begin{itemize}
\item data collection and feature engineering;
\item selection of characteristics for the model feature space;
\item clustering with different algorithms in searching for a model with the best quality metrics;
\item interpretation of the result, i.e. analysis of the difference between objects in different groups.
\end{itemize}

Further in the paper we successively consider the mentioned stages.

\section{Initial data}

In this section, we describe the databases and characteristics that have been used to compile the dataset of this project. Some of the characteristics are not directly used for the clustering because they are available for only a small number of blazars, nevertheless they are useful for subsequent analysis.

The basis for our dataset is the 5th edition of the Roma--BZCAT catalog of blazars \citep{2009A&A...495..691M,2015ApSS.357...75M}. The catalog contains a list of 3561 AGNs which are classified by the authors as blazars based on their observed properties. The 
following information is available: coordinates; redshifts; optical magnitudes in the $R$~band from USNO-B1.0, $r$~filter from SDSS~DR10, or in other filters when these data are absent; 1.4~GHz (NVSS, FIRST, 21~cm,) and 143~GHz (Planck, 2.1~mm)
radio flux densities; X-ray (ROSAT, Swift-XRT) and gamma-ray (Fermi-LAT) fluxes as well as the radio-to-optical spectral index characterizing the ratio between the radio and optical emission. 
Notice that the $R$~band magnitude presented in Roma-BZCAT
describes the optical radiation rather loosely, being sometimes obtained from different photometric filters. 
For this reason we used more consistent data on optical magnitudes from other catalogs.

Based on the BLcat\footnote{\tt https://www.sao.ru/blcat/} RATAN-600 measurements at frequencies 1--22~GHz covering the period of observations 2006--2022 \citep{2014A&A...572A..59M,2022AstBu..77..246S} and the CATS database\footnote{\tt https://www.sao.ru/cats/} \citep{1997BaltA...6..275V, 2005BSAO...58..118V}, we calculated the averaged spectral flux density at a frequency of 5~GHz, radio spectral indices, and radio variability.
The averaged spectral indices $\alpha$ were calculated for the 1--2, \mbox{2--5}, \mbox{5--8}, 5--11, \mbox{8--11}, 11--22, 8--22, and 5--22~GHz ranges. The radio variability is given at frequencies of 1, 2, 5, 8, 11, and 22~GHz. 
The variability index was calculated using the formula from \cite{1992ApJ...399...16A}:
\begin{equation}
V = \frac
{(S_{\max}-\sigma_{S_{\max}}) - (S_{\min}+\sigma_{S_{\min}})}
{(S_{\max}-\sigma_{S_{\max}}) + (S_{\min}+\sigma_{S_{\min}})},
\end{equation}
where $S_{\max}$, $S_{\min}$ are the maximum and minimum flux densities, and $\sigma_{S_{\max}}$, $\sigma_{S_{\min}}$ are their standard errors.

The infrared measurements are represented by data from the 
Wide-field Infrared Survey Explorer (WISE) in the W1, W2, W3, W4 bands (3.4, 4.6, 12, 22~{\textmu}m) and by the Two Micron All-Sky Survey (2MASS) data in the JHK bands (1.25, 1.65, 2.2 {\textmu}m). The data were taken from the NASA/IPAC Infrared Science Archive\!\footnote{\tt https://irsa.ipac.caltech.edu} using the {\tt pyvo} Python library,\!\footnote{\tt https://pypi.org/project/pyvo/} identification of the blazars was carried out by coordinates using the cone search query in the $9\arcsec$ field of view (the WISE angular resolution is about $6\arcsec$).

The optical range is represented by the \mbox{Pan-STARRS}\footnote{\tt https://outerspace.stsci.edu/display/PANSTARRS/} measurements in the $grizy$ filters (effective wavelengths of 4810, 6170, 7520, 8660, and 9620~{\AA}). The data was obtained by a standard request to the archive with a list of object coordinates. Additionally, for the optical range we downloaded the Sloan Digital Sky Survey (SDSS) DR17\footnote{\tt https://skyserver.sdss.org/dr17/} data in the $ugriz$ filters (effective wavelengths 3557, 4702, 6175, 7491, and 8946~{\AA}) via the provided web form (the requests were made automatically using a script).

The UV range is represented by the GALEX FUV and NUV channels (effective wavelengths of 1538.6 and 2315.7~{\AA}). The data was obtained from the Mikulski Archive for Space Telescopes (MAST) using the {\tt astroquery} Python library. A UV counterpart was identified as the object closest to the coordinates in the $7\farcs2$ field of view (GALEX angular resolution is $4''$).

In all the above cases, if there were several objects in the field of view of a search query, we chose the closest by angular distance. The possible presence of outliers was additionally controlled by histograms of angular distance deviations from the blazar coordinates. Since the identification was carried out in automatic mode, we cannot completely exclude incorrect identifications in some cases, but since we took the minimum possible search radius, comparable to the resolution of the instruments, such cases should be rare and can be discarded during subsequent data cleansing.

The extinction was determined using the NED's Coordinate and Galactic Extinction Calculator.\!\footnote{\tt http://ned.ipac.caltech.edu/forms/calculator.html} For the GALEX FUV and NUV channels, we used the extinction law from \cite{1999PASP..111...63F}, the calculations were made with the {\tt extinction} Python library.\!\footnote{\tt https://extinction.readthedocs.io/en/latest/} For the WISE IR range, the extinction was considered zero.

To determine the peak frequency of the synchrotron component, we used spectral energy distributions (SEDs) obtained from the SED Builder\footnote{\tt https://tools.ssdc.asi.it/SED/} tool of the Italian Space Agency (ASI) Space Science Data Center (SSDC). The measurements were downloaded using Selenium WebDriver,\!\footnote{\tt https://www.selenium.dev} which allows interaction with web sites in an automated mode.

Thus, in the initial dataset we managed to collect a fairly extensive set of observed data in various ranges of the electromagnetic spectrum: from radio to gamma emission. The dataset also includes information about redshift, spectral indices, and estimates of variability in the radio range. In the next section, we describe additional processing of the derived data in order to extract more informative features and present a complete dataset of the obtained and calculated characteristics.

\section{Calculation of blazars' characteristics}
\label{sec:character_calc}

The initial characteristics obtained from catalogs often cannot be directly used in the model, because they might describe not only the properties of the objects but also other factors affecting the result. For example, the magnitudes depend on the photometric system of a particular catalog. Therefore, to solve our problem, we should obtain such characteristics that in the best possible way describe the physical properties of the blazars.

Since blazars are located at different cosmological distances, these  characteristics should be related to the rest frame of an object. Unfortunately, this is often impossible in practice. For instance, to estimate the luminosity at a certain frequency at cosmological distances, a good description of the SED shape is necessary, but for most of the blazars we have only point estimates of this shape at a number of frequencies; moreover, the SED can be variable. Empirical analytical dependencies (see, e.g., \citealt{2010MNRAS.405.1409C}) work only at small redshifts $z<0.5$.

Here we will use the rest frame characteristics where possible, the remaining features will be given in the observer's frame of reference, and the distance to a blazar will also be set as one of the parameters. In Section~\ref{sec:feature_space} we describe how this might affect the results in more detail.

\begin{figure*}
\centering
\includegraphics[width=\textwidth]{
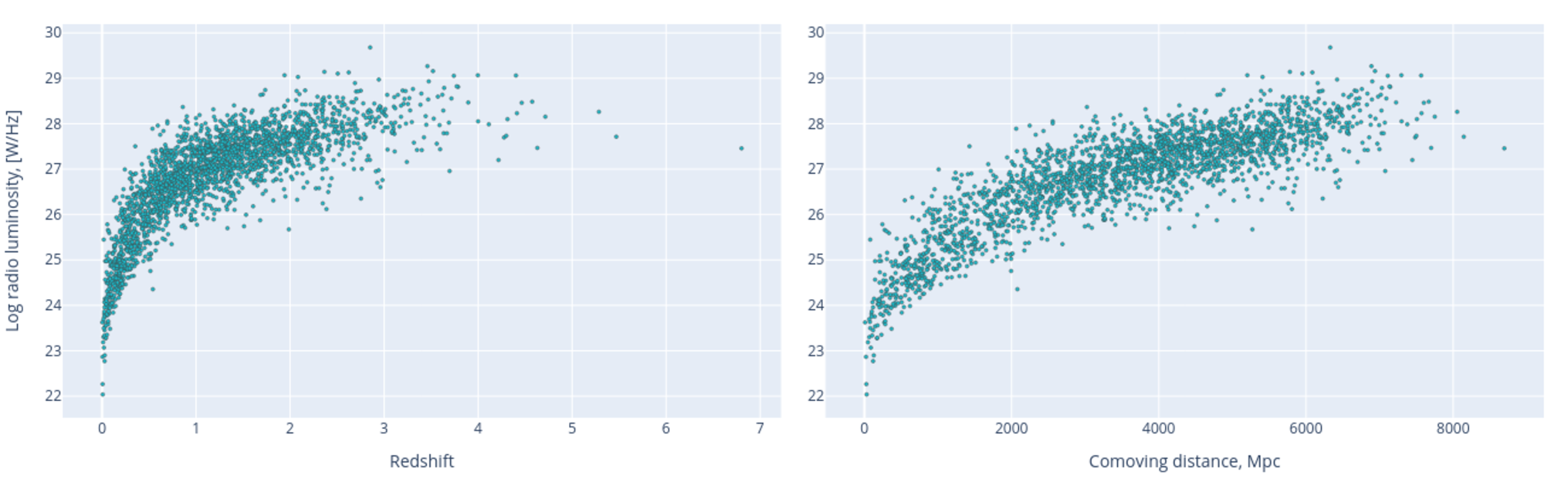
}
\caption{Logarithm of radio luminosity (at 5~GHz) vs.\ redshift (left) and comoving distance (right).}
\label{fig:RLum_Dist}
\end{figure*}

The distance to a blazar could be described directly by the redshift, but in this scale the distance distribution of the blazars appears rather crowded and uneven. In Fig.~\ref{fig:RLum_Dist} the dependences of the monochromatic radio luminosity on the redshift and on the comoving distance are presented. The comoving distance scale is more suitable for the modeling, the distances were determined from the redshifts using the {\tt astropy} library \citep{2022ApJ...935..167A} and based on the $\Lambda$CDM cosmology with the Planck Collaboration parameters \citep{2016AA...594A..13P}: $H_0=67.74$~km\,s$^{-1}$\,Mpc$^{-1}$, $\Omega_m=0.3089$, $\Omega_\Lambda=0.6911$. Along with this, some other parameters were calculated: luminosity distance, distance modulus, lookback distance, and the Universe's age in the blazar rest frame at the time of light emission. The monochromatic (5~GHz) radio luminosity was estimated by the formula:
\begin{equation}
L_{5} = 4 \pi D^2_L S_{5} (1+z)^{-\alpha-1},
\end{equation}
where $D_L$ is the luminosity distance, $S_{5}$ is the flux density at 5~GHz, $z$ is the redshift, $\alpha$ is the averaged spectral index taken for 5--11~GHz or, where the measurements were absent, for 5--8~GHz.

The dependence of radio luminosity on distance in Fig.~\ref{fig:RLum_Dist} is completely or partially caused by the selection effects. One of them is the Malmquist bias (e.g., \citealt{2005MNRAS.362..321B}): roughly speaking, if we assume a normal luminosity distribution, the same at all distances $d$, then with increasing distance the detection limit shifts to higher luminosities, which reduces the number of objects in the left wing of the distribution, and simultaneously the expected number of high-luminosity objects grows with increasing area of the sphere of radius $d$, which increases the probability of detecting bright objects in the right wing of the distribution. Since both effects are proportional to $d^2$, a linear increase in average luminosity with distance should be observed. It must be noted that the observed dependence is not linear at distances $\lesssim\!2000$~Mpc; however, selection cannot be excluded in this case either, since the measurements were obtained in different surveys, and the interest to the nearest but not necessarily luminous objects is natural.

The optical flux densities in the observer's frame of reference were calculated based on the AB magnitude system as
\begin{equation}
\log_{10} (\nu F_{\nu}) = \log_{10} (\nu F_{\nu 0}) - \frac{m_{\nu}-E_{\nu}}{2.5},
\end{equation}
where $\nu$ is the effective frequency of a photometric band, $F_{\nu 0}$ is the flux density from zero magnitude, $m_{\nu}$ and $E_\nu$ are the magnitude and extinction in the photometric band (extinction according to the NED data). The $\nu$ and $F_{\nu 0}$ values were taken from the descriptions of corresponding photometric systems \citep{2007ApJS..173..682M, 2012ApJ...750...99T, 2003AJ....126.1090C, 2011ApJ...735..112J}. For SDSS we implemented the corrections $u = u - 0.04$, $z = z + 0.02$, according to accepted practice.
Summary data are given in Table~\ref{tab:m2flux}.

\begin{table}
\caption{Parameters of the magnitude-to-flux-density transformations}
\label{tab:m2flux}
\small
\begin{tabular}{lcc}
\hline
Band & Effective wavelength & Zero magnitude \\
     &      ($\lambda = c/\nu$)         & flux density $F_{\nu 0}$, Jy \\
\hline
GALEX FUV    & 1538.6~\AA         & 3631 \\
GALEX NUV    & 2315.7~\AA         & 3631 \\
SDSS  u      & 3557~\AA           & 3631 \\
SDSS  g      & 4702~\AA           & 3631 \\
SDSS  r      & 6175~\AA           & 3631 \\
SDSS  i      & 7491~\AA           & 3631 \\
SDSS  z      & 8946~\AA           & 3631 \\
Pan-STARRS g & 4810~\AA           & 3631 \\
Pan-STARRS r & 6170~\AA           & 3631 \\
Pan-STARRS i & 7520~\AA           & 3631 \\
Pan-STARRS z & 8660~\AA           & 3631 \\
Pan-STARRS y & 9620~\AA           & 3631 \\
2MASS J      & 1.235~{\textmu}m   & 1594 \\
2MASS H      & 1.662~{\textmu}m   & 1024 \\
2MASS K      & 2.159~{\textmu}m   & 666.8 \\
WISE W1      & 3.3526~{\textmu}m  & 309.54 \\
WISE W2      & 4.6028~{\textmu}m  & 171.787 \\
WISE W3      & 11.5608~{\textmu}m & 31.674 \\
WISE W4      & 22.0883~{\textmu}m & 8.363 \\
\hline
\end{tabular}
\end{table}

\begin{figure}
\centering
\includegraphics[width=\columnwidth]{
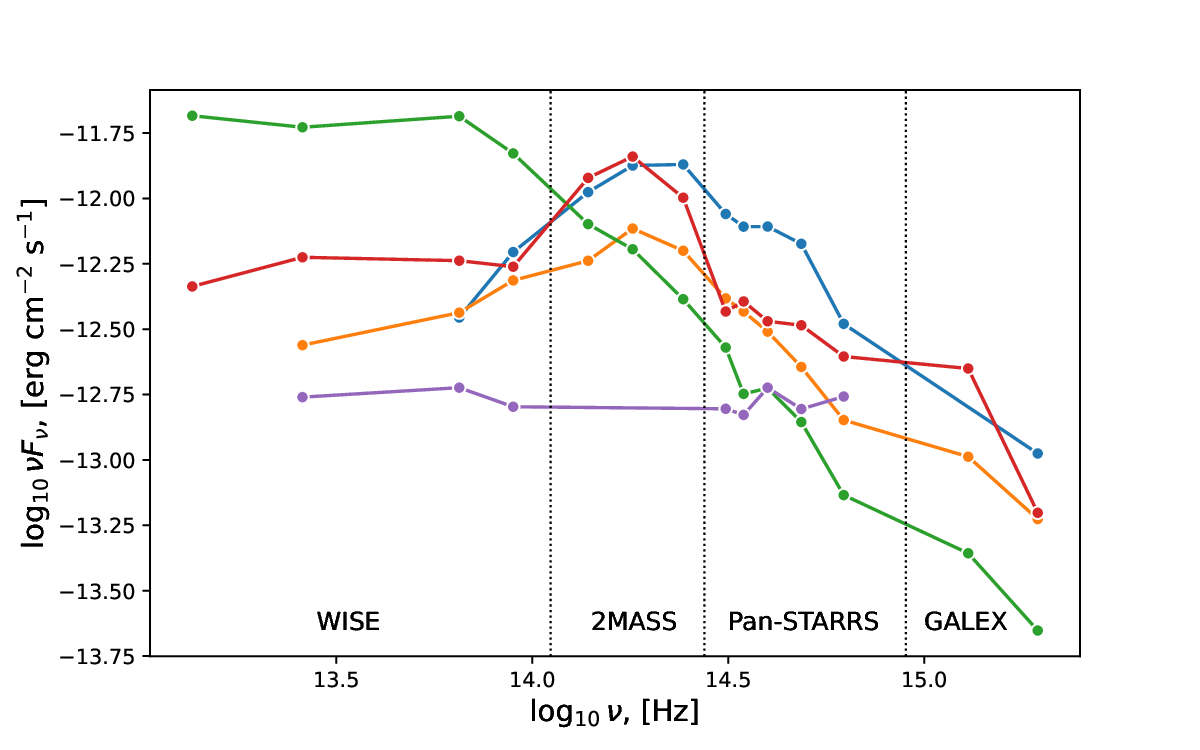
}
\caption{Spectral energy distributions 
for 5 randomly selected blazars, constructed 
based on the data from the WISE, 2MASS, Pan-STARRS, and GALEX catalogues.}
\label{fig:opt_fluxes}
\end{figure}

Based on the available stellar magnitudes and extinctions, we calculated optical colors in various photometric systems. However, their use in the clustering seems impractical. In Fig.~\ref{fig:opt_fluxes}, SEDs for five random blazars are presented, the points mark flux densities $\log_{10}\nu F_{\nu}$ in the WISE, 2MASS, Pan-STARRS, and GALEX passbands, for better visualization of individual SEDs the points are connected by lines. Flux densities are related to stellar magnitudes, and the difference between the pairs of the points generally represent the optical colors. It can be seen that in the frequency range of each individual instrument, for example WISE and Pan-STARRS whose data we further use in the clustering, a predominant slope of the spectrum can be assigned for a particular blazar, while the ratios between flux densities for the pairs of points (``colors'') can sometimes differ significantly. Thus, the use of colors in the clustering can create additional noise that would not allow the algorithm to estimate the predominant slope of the spectrum. For this reason, we calculated special features: tangents of the spectrum slope in the WISE and Pan-STARRS ranges (the slopes for 2MASS and GALEX were not used due to lack of data).
The spectrum slopes were approximated by a linear dependence
using the method of least squares.

\begin{figure*}
\centering
\includegraphics[width=\textwidth]{
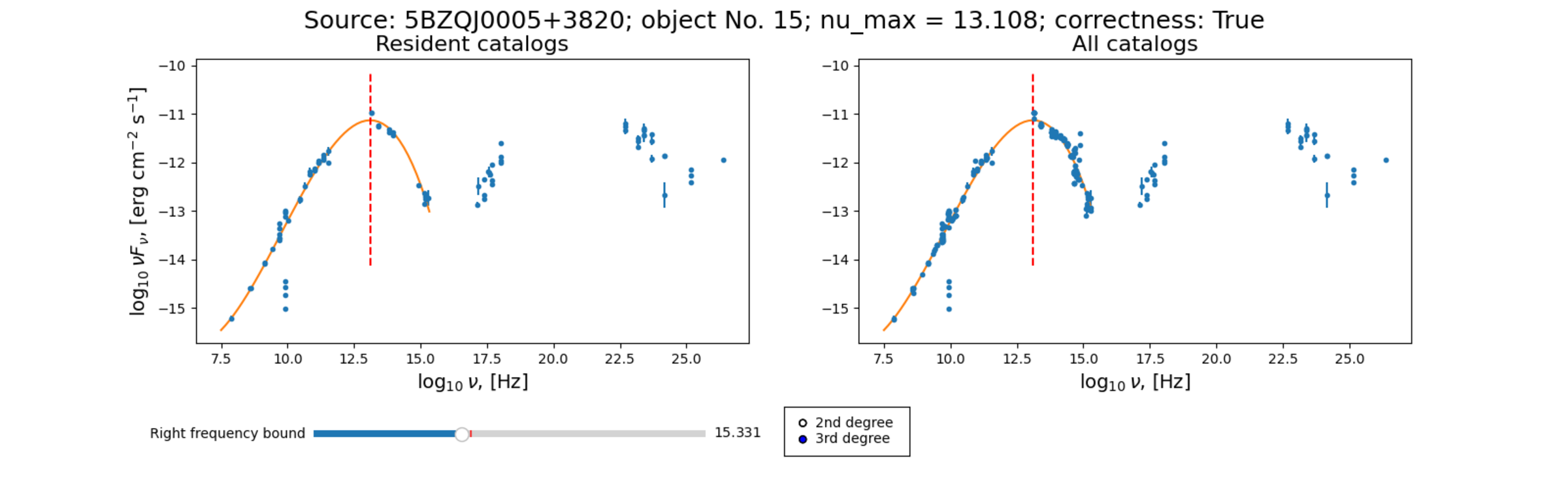
}
\caption{An example of SED fitting with a cubic polynomial to find the peak of the synchrotron component.}
\label{fig:sed_poly}
\end{figure*}

\begin{figure}
\centering
\includegraphics[width=\columnwidth]{
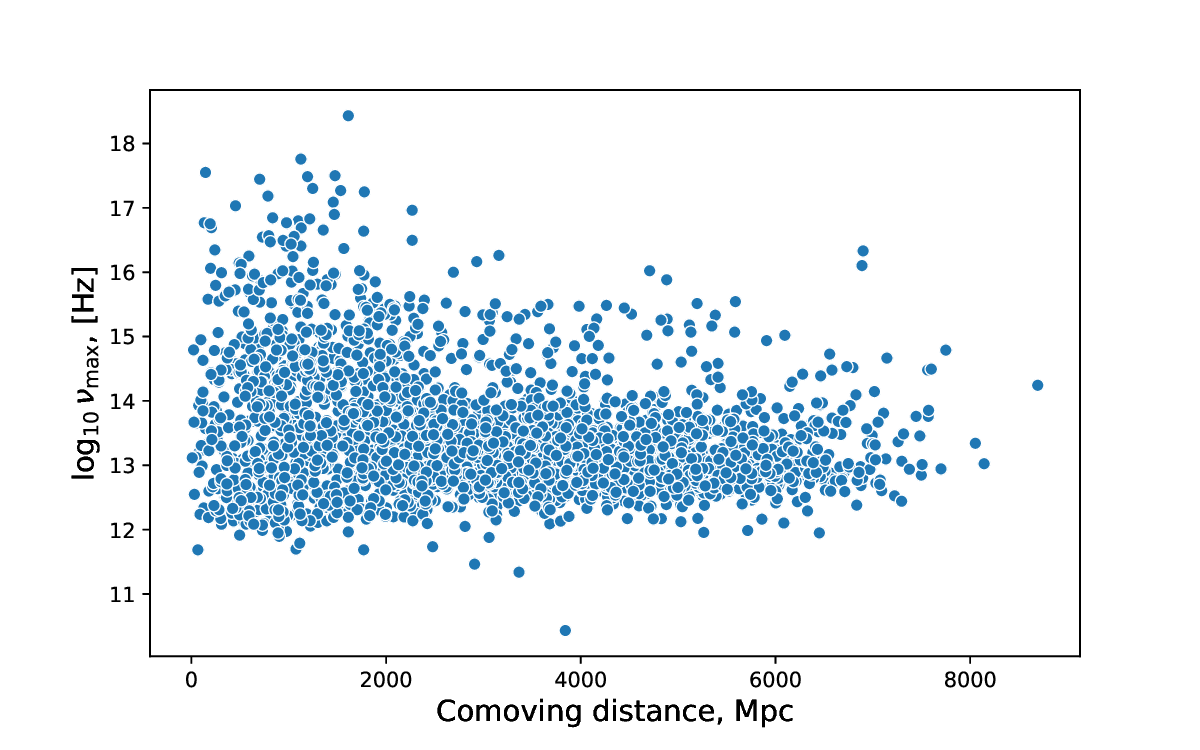
}
\caption{Frequencies of synchrotron peaks depending on comoving distance.}
\label{fig:synchro_max}
\end{figure}

To determine the frequency of the synchrotron peak, we used flux densities downloaded from the ASI Space Science Data Center SED Builder.
The position of the peak was determined by approximating a SED with polynomials of the third or, in some cases, second degree.\!\footnote{https://github.com/DKudryavtsev/BZCAT-SED-Viewer}
The flux density measurements were represented by the data from the Space Science Data Center resident catalogs and other catalogs. To calculate the parameters of the polynomial, the program used only the resident catalogs, while we visually controlled the obtained result using all the measurements (see Fig.~\ref{fig:sed_poly}).
The frequencies were transformed to the source's rest frame:
\begin{equation}
\nu_{\rm peak} = \nu_{\rm peak, obs} (1+z)
\end{equation}
A graph of the obtained values versus comoving distance is shown in Fig.~\ref{fig:synchro_max}.

\begin{figure}
\centering
\includegraphics[width=\columnwidth]{
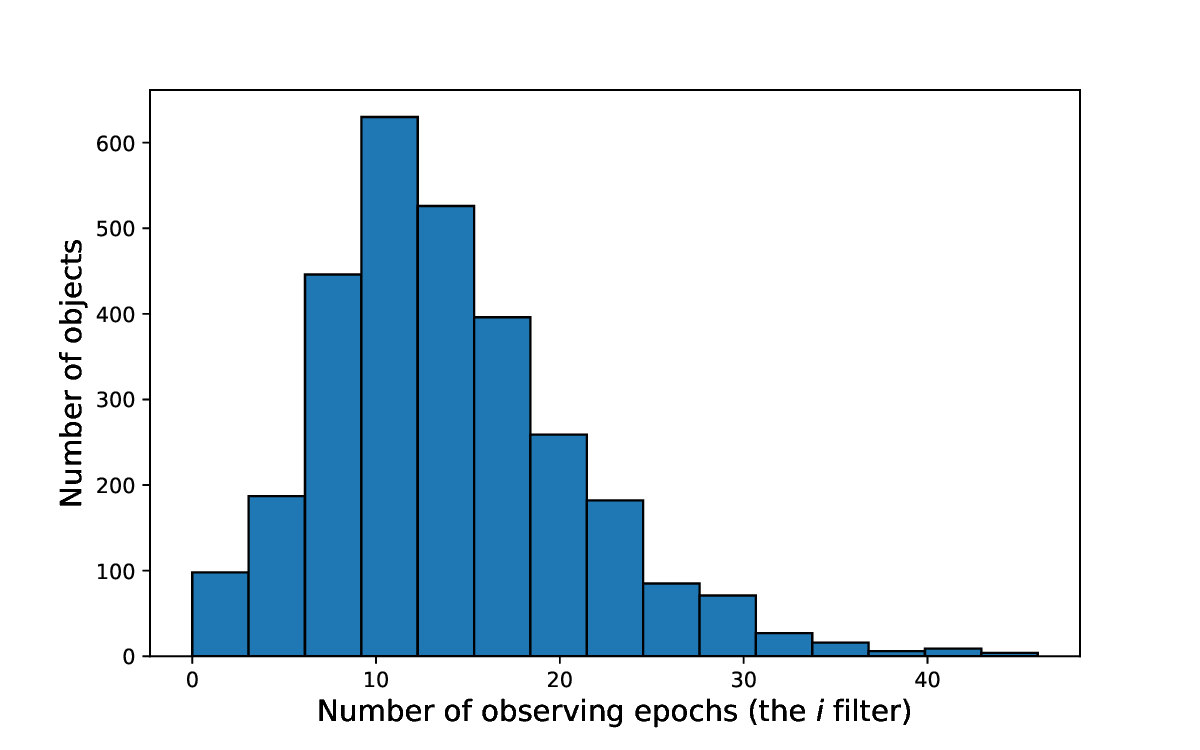
}
\caption{Distribution of the number of observing epochs for the Pan-STARRS $i$ filter (``iMeanPSFMagNpt'' in the catalogue).}
\label{fig:pstarrs_nobs}
\end{figure}

The estimates of optical variability were calculated 
from the minimum and maximum point-spread function (PSF) magnitudes presented in the Pan-STARRS data for each of the 5 filters ($grizy$). Our variability estimates are simple differences between
these values for all the blazars with two or more observing epochs (according to the number of the measurements included in the mean PSF magnitude from the detections in a corresponding filter). The estimates are rough, as they depend on the time when the observing epochs have been carried out and on their number. As an example, the distribution of the number of observations in the $i$ filter is shown in Fig.~\ref{fig:pstarrs_nobs}.

Roma-BZCAT presents X-ray fluxes for the range \mbox{0.1--2.4}~keV (5--124~\AA), we recalculated them into the logarithmic scale $\log_{10}$[erg\,cm$^{-2}$\,s$^{-1}$]. Gamma radiation is represented in Roma-BZCAT in photons\,cm$^{-2}$\,s$^{-1}$ for the range 1--100~GeV, we also recalculated the values into the scale $\log_{10}$[erg\,cm$^{-2}$\,s$^{-1}$], taking the middle of the range as the photon energy: 50~GeV.

In addition to the Roma-BZCAT radio-to-optical spectral index, which characterizes the ratio between the radio and optical fluxes, we calculated other parameters that describe flux ratios at different frequencies of the electromagnetic spectrum (let us call these parameters the ``hardnesses''). They are calculated as decimal logarithms of the ratios between flux densities at the frequencies $\nu F_{\rm 1.4 GHz}$ (radio), $\nu F_{\rm W2}$ (IR), $\nu F_{i}$ (optics), $\nu F_{\rm X}$ (X-rays), and $\nu F_{\gamma}$ (gamma rays); e.g., the IR/optics hardness is $\log_{10}(\nu F_{\rm W2}\,/\,\nu F_{i}) \equiv \log_{10}(\nu F_{\rm W2}) - \log_{10}(\nu F_{i})$. The clustering model uses six such ratios, because not enough data is available for all the frequencies.

\begin{figure*}
\centering
\includegraphics[width=0.95\textwidth]{
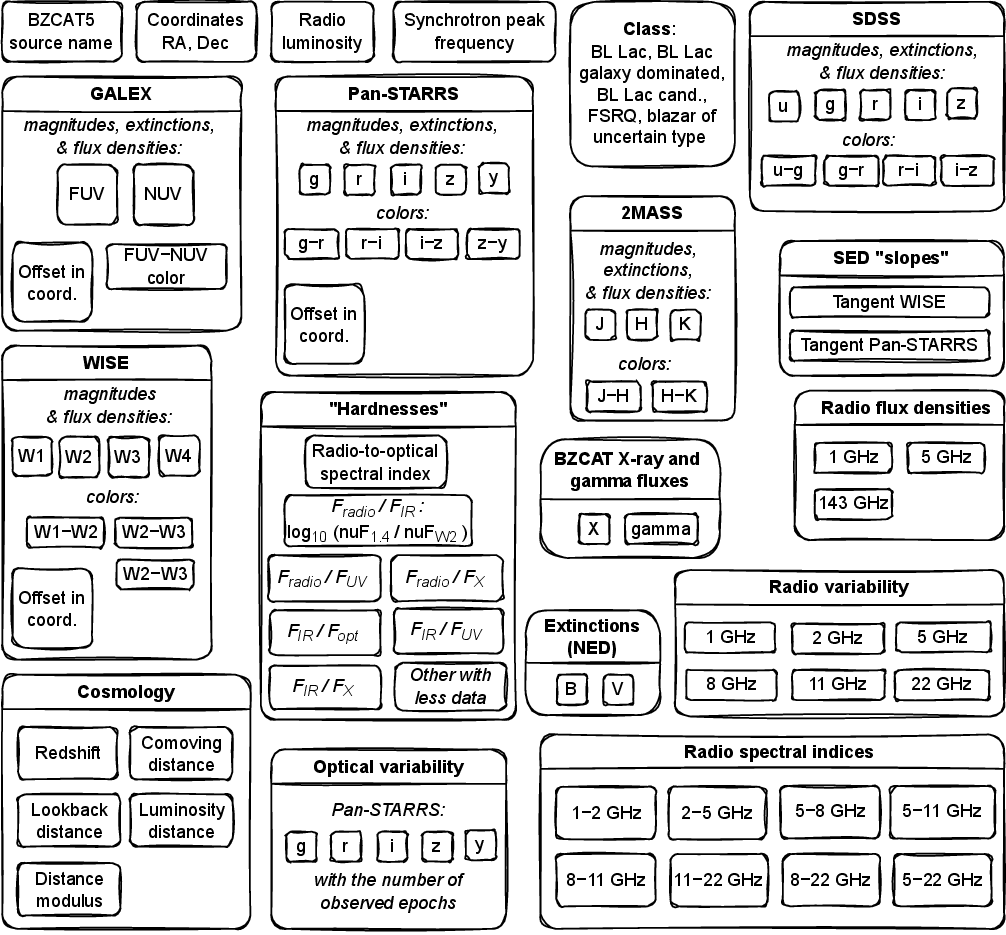
}
\caption{All dataset characteristics. The data are 
presented in the VizieR database (CDS).}
\label{fig:features}
\end{figure*}

The complete list of the parameters available in the final dataset includes more than 100 items. The dataset is schematically presented in Fig.~\ref{fig:features} and is 
available at CDS.\!\footnote{https://cdsarc.cds.unistra.fr/viz-bin/cat/J/other/RAA}

\section{Feature space and model dataset}
\label{sec:feature_space}

The immediate use of an entire dataset by ML algorithms is impossible. First of all, the data must be appropriately preprocessed to obtain meaningful results. Moreover, some features may have a large amount of missing data, some strongly correlate with each other, others are auxiliary and are not related to the actual properties of the objects (e.g., extinction), and some should be discarded since they do not affect the final result but increase the dimensionality. The best practice also is when we could provide the ML model not only with the direct characteristics of the objects but also with ML-specific features, which are combinations or transformations of actual characteristics that help the model to ``understand'' data better.  Therefore, the next step after collecting the data is the construction of the model dataset that will be directly used by the ML algorithms. This includes the selection of characteristics relevant to the problem, feature engineering and transformations, data cleansing, imputation of missing values, scaling, etc. 

Notice that while this model dataset is constructed to be used by ML algorithms and undergoes certain transformations during this process, the predictions that a trained model produces in the end, such as the cluster label in our case, is object-specific. In other words, having cluster labels (membership) for the blazars as a result of our clustering, we can further analyze any other characteristics of the blazars from the original or model dataset along with even the new ones.

The choice of the features to form the feature space of the clustering can be made using various approaches. In our case we used as many available blazar characteristics as possible. In general, if there is no a predetermined scope of investigation, i.e., the study is not aimed at revealing relationships between some specific characteristics selected beforehand, this approach allows us to describe the objects under investigation most completely, form the groups of similar objects without a priori assumptions, and increase the reproducibility of the clustering results.

Here we consider in detail the preparation of the model dataset to be directly used by the clustering algorithms.

\subsection{Dropping unnecessary characteristics}

Different ways of describing cosmological distances (the ``Cosmology'' cell in Fig.~\ref{fig:features}) are a priori related by analytical dependencies and do not give new information to the model, therefore
only one of them, the comoving distance, was chosen, as the distribution of blazars in this scale is most uniform (see Fig.~\ref{fig:RLum_Dist} above).

Stellar magnitudes and flux densities are also analytically related. For the model dataset the latter were selected since they do not depend
on the photometric system and more directly connected to physical properties: the luminosity of an object and the distance to it.

The RA, Dec coordinates were excluded as we do not expect heterogeneity here and also because the spherical coordinate system (full circle in RA, semicircle in Dec) leads to an artificial global structure in the data.

The blazar types according to the Roma-BZCAT classification (BL\,Lacs, FSRQs, etc.) is a categorical feature, which in combination with other characteristics having continuous distributions leads to a trivial solution: division into the known types. Therefore, this information was removed from the model dataset.

As noted above, instead of the colors we used more smoothed parameters: tangents of the spectrum slope in the WISE and Pan-STARRS passbands. For this reason, colors were not considered in modeling. Other photometric passbands were not included due to lack of data (see below).

Unfortunately, we had to exclude from the model dataset the data on radio and optical variability: these values significantly correlate with the number of observations \citep{2000AJ....120.2278T,2007AJ....133.1947N,2024arXiv240202283K}, which in our case generated an artificial cluster of radio variability: the differences were clearly visible at the most observed frequency of 5~GHz, while at other frequencies with fewer measurements the cluster was not that distinguished.

We also excluded the radio spectral indices. The modeling showed that their distributions for the groups found in clustering had not had significant differences, therefore the final model was built without them.

Extinctions were dropped as these are auxiliary data used in flux density calculations.

\subsection{Dropping characteristics with many missing values}

A separate problem is missing values: almost all characteristics, to a greater or lesser extent, are subjected to lack of data. The processing of missing values is covered further in more detail, but characteristics with a very large number of absent measurements cannot be used in modeling. According to accepted empirical practice, we excluded features with more than 40\% of missing data:
in particular, the GALEX FUV values and associated characteristics, SDSS and 2MASS data, Roma-BZCAT (Fermi) data on gamma-ray fluxes, and data on the radio flux densities at 143~GHz (1.4 and 5~GHz have remained).

We once again notice that characteristics excluded from the modeling can be used for analysis after the clustering.

\subsection{Outliers}
\label{sect:outliers}

Measurements outstanding significantly from the distribution of a characteristic can distort results of most clustering algorithms. We considered the distributions of the features selected for the modeling and visually evaluated their boundaries. Blazars outside designated distribution boundaries were excluded from the model dataset
(not more than several objects for a feature, a total of 34 objects), their classification into groups was carried out after the clustering using a separately trained k-nearest neighbors (KNN) model (section~\ref{sect:outliers_class}).

\subsection{Multicollinearity. Combining similar characteristics into meta-features}

The initial data contains sets of characteristics that naturally correlate with each other: the flux densities at different frequencies and part of the flux density ratios (hardnesses). Standard techniques in processing multicollinear features in machine learning are either their exclusion or the Principal Component Analysis (PCA): transformation of the features by linear algebra methods into new mutually orthogonal (zero correlation) characteristics oriented in the feature space along the axes of the greatest variance.

In this paper we used PCA to combine a number of flux densities at different frequencies into one meta-feature, using for that the first principal component. The sets of input characteristics and their corresponding meta-features are shown in Table~\ref{tab:meta_features}. The choice of these two meta-features is based on the simple core--jet model of AGNs, where the radio emission is unambiguously related to the synchrotron radiation from the jet, while emission in other electromagnetic ranges can be generated by both the core regions and the jet.

\begin{table}
\caption{Sets of similar (correlated) physical characteristics and their meta-features in the model dataset}
\label{tab:meta_features}
\small
\begin{tabular}{p{51mm}p{22mm}}
\hline
Characteristics  & Meta-feature \\
 \hline
 X-ray, GALEX NUV, Pan-STARRS $grizy$, and WISE W1--W4 flux densities &  
         ``Short-wavelength'' flux, $f_{\rm IR-X}$ \\[5pt]
Flux densities at 1.4 and 5~GHz & Radio flux, $f_{\rm radio}$  \\
\hline
\end{tabular}
\end{table}

Notice that for an individual blazar, measurements for some input characteristics may be missing. In such cases we imputed the missing values using probabilistic PCA (pPCA, \citealt{DBLP:journals/neco/TippingB99}). 
The pPCA implementation\footnote{\tt https://github.com/el-hult/pyppca} from \cite{porta:inria-00321476} was adopted.
The method is considered in more detail in Section~\ref{sec:imputation}.
We applied pPCA separately for each set of characteristics from Table~\ref{tab:meta_features}. 
If all the corresponding values for an object were missing, we left them empty (at this first stage).

In contrast to the above mentioned, for the hardnesses, some of which may also correlate with each other, it is important to preserve the information about differences in flux densities at various ranges of the electromagnetic spectrum. In this case, the multicollinearity was 
removed later during dimensionality reduction of the entire model dataset, also using PCA but taking more principal components (see Section~\ref{sec:classical_ml}).

\subsection{Scaling}

To ensure equal priority of features in PCA and clustering, 
all of them must be expressed in a unified numerical scale.
At all the stages where that was necessary, we used the scikit-learn \citep{JMLR:v12:pedregosa11a} standard scaler, which produces the zero mean and unit variance for a feature.

\subsection{Model dataset}

\begin{figure}
\centering
\includegraphics[width=\columnwidth]{
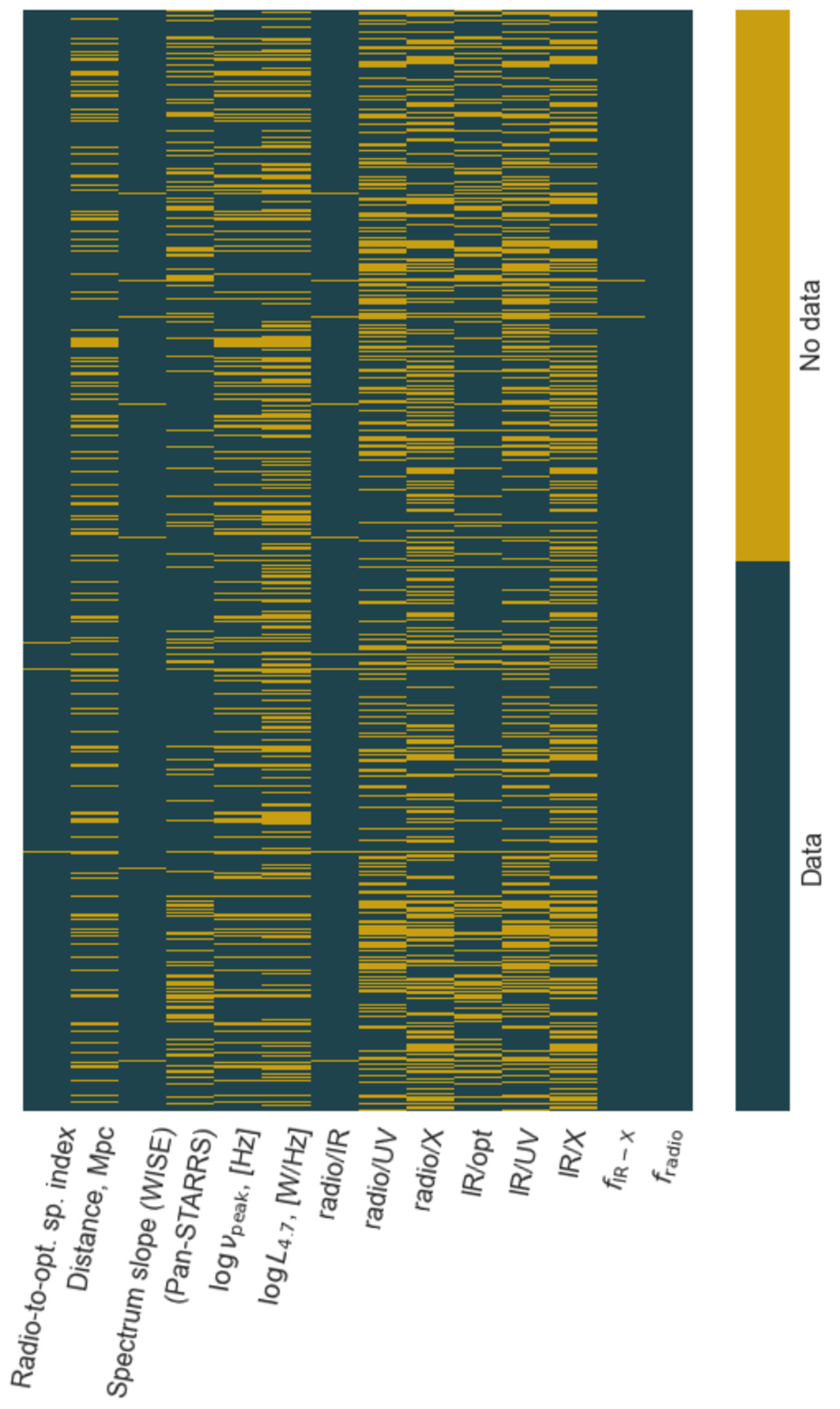
}
\caption{Model dataset heatmap. The rows ($y$-axis) correspond to individual objects. The columns ($x$-axis) are the features. Missing data are shown by the light color.}
\label{fig:model_dataset}
\end{figure}

The clustering model dataset after data cleansing and feature transformations includes 14 features. The heatmap of the dataset is shown in Fig.~\ref{fig:model_dataset}. The columns of the table are designated along the $x$ axis, the $y$ axis corresponds to its rows, in which individual object vectors are located. The heatmap shows missing data (less than 40\%\ in each of the columns).

\section{Selection and other hindering~effects}

We should notice that our feature space is subjected to some effects that are negative for interpretation of the results that could be obtained from the clustering. In the first place, all selection effects are preserved, and almost all characteristics are dependent on the distance to the blazars (or redshift~$z$). For example, as has already been mentioned in Section~\ref{sec:character_calc}, the redshift-corrected radio luminosity nevertheless shows strong dependence on $z$ due to the Malmquist effect. Even the flux density ratios are dependent on the distance because of the cosmological rest frame drift, which could not be corrected due to the absence of an accurate SED model for each of the blazars. 

At the same time, these effects can be considered useful for the clustering because they could potentially help separate classes that naturally demonstrate different distance distributions (because of the selection in the data or not); they also contribute to a more accurate probabilistic imputation of missing values (see Section~\ref{sec:imputation}). It is for these reasons that we leave the comoving distance and raw flux densities as model characteristics. 
The dependence on distance must be kept in mind during further analysis of the obtained groups. 

The second nuisance is the fact that blazars are variable sources. In our dataset we took the average characteristics of the objects in the way they are presented in most catalogs. In some cases different characteristics may be measured in different states of blazar activity (active/quiescent) (see, e.g., \citealt{2014MNRAS.442..629R}). This restricts our results to only the groups' statistical properties, any conclusion for an individual source must be treated with great caution.

Finally, the BZCAT catalog is not a complete flux-limited list of blazars. Although the incompleteness of the sample still allows us to perform the clustering and analyze the observed differences, it could influence the distribution of blazars within the clusters, i.e., population of certain groups (boundaries of the clusters in the feature space) may change for a more complete sample. 
We evaluate the effect of data incompletness in more detail in Section~\ref{sec:robustness}.

\section{Clustering}

The clustering was carried out first with a subsample of blazars that had no missing data in the model dataset (858 blazars, $\sim\!\!24$\%\ of the BZCAT catalog) and then with the full sample and imputed missing data. As well, we tested various clustering algorithms: several from the scikit-learn library \citep{JMLR:v12:pedregosa11a} and Kohonen's self-organizing maps (SOMs, \citealt{2001som..book.....K, JSSv078i09}) based on a competitive neural network.

\subsection{The subsample without missing values}

\subsubsection{Clustering with k-means and PCA~dimensionality~reduction}
\label{sec:classical_ml}

We experimented with several clustering methods offered by the {\tt scikit-learn} library \citep{JMLR:v12:pedregosa11a}: k-means, Gaussian mixture, agglomerative clustering, and spectral clustering.
The results were compared by the internal clustering validation metrics: the silhouette \citep{Rousseeuw:1987:SGA:38768.38772}, Calinski--Harabasz \citep{calinski1974}, and Davies--Bouldin \citep{davies_cluster_1979} scores. As the final option with the best indicators, the combination of PCA dimensionality reduction with k-means clustering was chosen. For dimensionality reduction, we took an explained variance of 90\%\ as a criterion, thus converting the 14 model dataset features into 6 metafeatures: mutually orthogonal (uncorrelated) principal components. After that, k-means clustering \citep{1283494} was performed in this 6D space.

A comparison of clustering validation metrics calculated for various number of clusters shows that the data distribution is, not surprisingly, a continuous cloud without localized groups. Thus, the number of clusters may be determined pretty loosely.
We selected the number of clusters based on the best ($\sim\!90\%$) match of the clustering results for the subsample without missing values and for the full sample, see Section~\ref{sec:imputation}.
With this approach, the optimal number of clusters turns out to be five.
For comparison, the popular ``elbow'' method, taken as the first approximation and based on the analysis of decreasing distortion
(average squared Euclidean distance from the centroids of the respective clusters), gives the number of clusters equal to four.

These results are further used as a baseline model for clustering the entire Roma-BZCAT catalog, and are also compared with the approach based on self-organizing maps. The visualization of the obtained clusters is presented in Section~\ref{sec:comparison} along with the self-organizing maps.

\begin{figure}
\centering
\includegraphics[width=\columnwidth]{
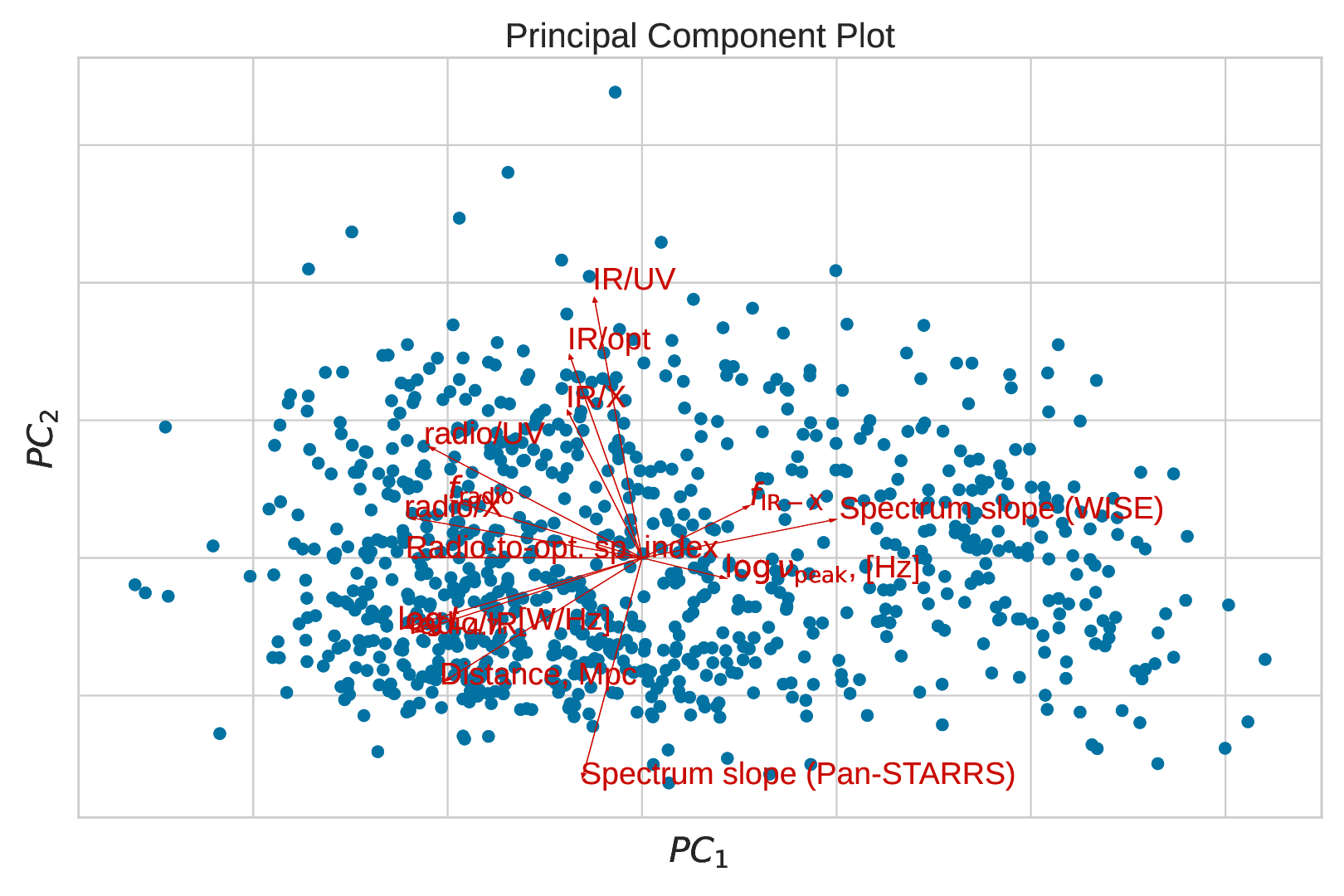
}
\caption{The PCA biplot with the projections of the dataset onto the first two primary components. The vector lengths correspond to the importance of the features for the clustering result.}
\label{fig:biplot}
\end{figure}

The PCA also allows us to estimate the significance of the features for the clustering. To this end, we constructed a PCA biplot (Fig.~\ref{fig:biplot}) using the {\tt Yellowbrick} library \citep{Bengfort2019}. The figure shows the projection of the dataset onto the plane of the two primary components, and the lengths of the vectors correspond to the importance of each feature, reflected by the magnitude of the corresponding values in the eigenvectors of the primary components. The directions of the vectors also demonstrate the degree of correlation (same direction) or anticorrelation (opposite direction) between the features.
Numerically, the contributions are given in Table~\ref{tab:importance}.
Notice that here we do not use the characteristics related to gamma-ray measurements because of the scarcity of data. The gamma-ray range, nevertheless, may be of great importance for blazar classification. We consider this problem in more detail in Section~\ref{sec:robustness} and give there a similar table (Table~\ref{tab:importance_gamma}), where the importance of the features is recalculated taking into account the gamma-ray emission.

\begin{table}
\centering
\caption{Importance of the features (without gamma-ray range)}
\label{tab:importance}
\small
\begin{tabular}{lc}
\hline
Feature & Contribution, \% \\
\hline
IR/UV                       & 10.9 \\
Spectrum slope (Pan-STARRS) & 9.3 \\
IR/opt                      & 8.7 \\
Radio/UV                    & 8.0 \\
Distance                    & 7.9 \\
$\log L_{5}$              & 7.9 \\
Radio/IR                    & 7.6 \\
Radio/X                     & 7.3 \\
Radio-to-opt. sp. index     & 7.1 \\
IR/X                        & 6.5 \\
$f_{\rm radio}$             & 6.3 \\
Spectrum slope (WISE)       & 6.1 \\
$f_{\rm IR-X}$              & 3.8 \\
$\log \nu_{\rm peak}$       & 2.6 \\
\hline
Total                       & 100 \\
\hline
\end{tabular}
\end{table}

\subsubsection{Self-organizing maps}

Kohonen's self-organizing maps (SOMs; \citealt{2001som..book.....K})
are a neural network with competitive learning, used for clustering and
visualization of multiparametric data that can contain non-obvious
patterns.
In particular, SOMs solve the problem of projecting a multidimensional
space into lower dimensions: onto a plane or into a 3D
space, where data vectors are grouped according to the degree of similarity of their parameters, which allows one to perform the clustering, e.g., to separate different populations
of sources.

There are many software packages for data analysis by the SOM method, written in different programming languages. In this study we chose a Python package {\tt somoclu}\footnote{ \url{https://github.com/peter wittek/somoclu}} \citep{JSSv078i09}.

In the SOM clustering, a grid of $200\times320$ output neurons was built with the number of weights for each neuron equal to the dimension of the input vector (our object). The SOM algorithm finds the Euclidean distance 
between the vectors in a multiparametric space (the parameters are scaled to the interval [0, 1]) and adjusts the weights of the neurons so that they would be structurally similar to the distribution of the input vectors in the feature space. In this way the input vectors (objects) become arranged on certain areas on the output 2D SOM map in such a way that objects with similar parameters are located close to each other. At the same time, the distribution of the neuron weight vectors in the feature space becomes close to the data distribution. In other words, after training the network we have an ordered 2D structure of neurons with the high-dimensional data topology encoded in their multidimensional weights. 

The final step is also k-means, but in this case we make it using the weights of the trained neurons and then labelling the objects according to the cluster label of their nearest neuron. The advantage of this method over the PCA dimensionality reduction is that it can restore possible nonlinearities in data distribution, while the PCA is a more straightforward and interpretable method of linear algebra.

\subsubsection{Cluster visualization and comparison of the two methods}
\label{sec:comparison}

\begin{figure*}
\centering
\includegraphics[width=\textwidth]{
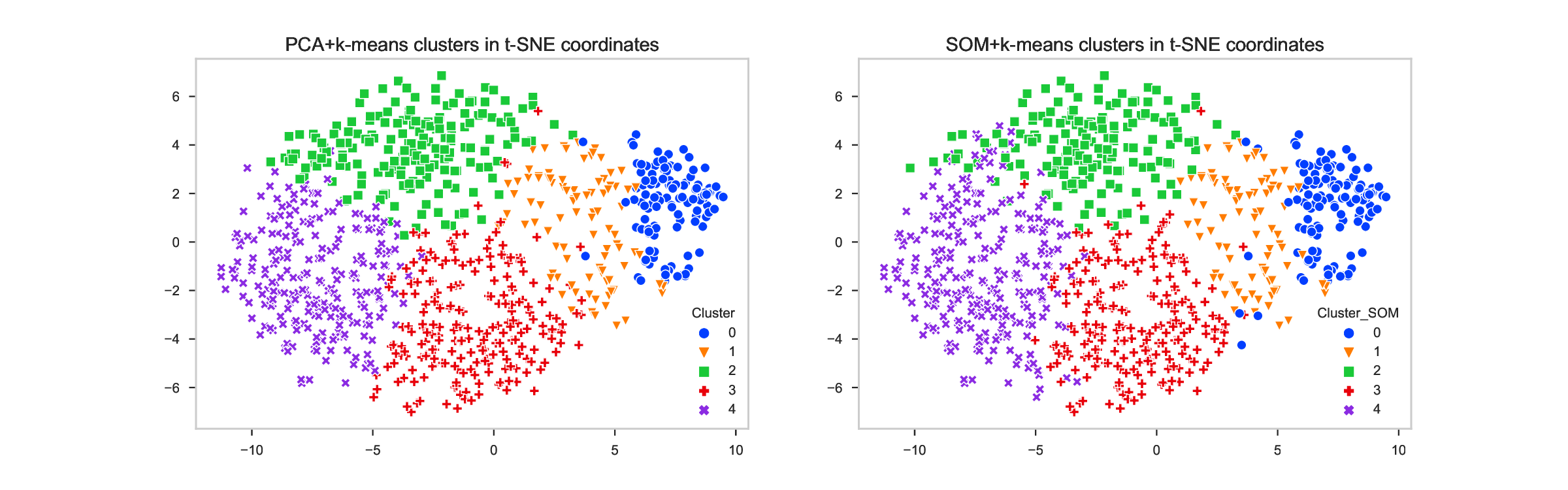
}
\includegraphics[width=\textwidth]{
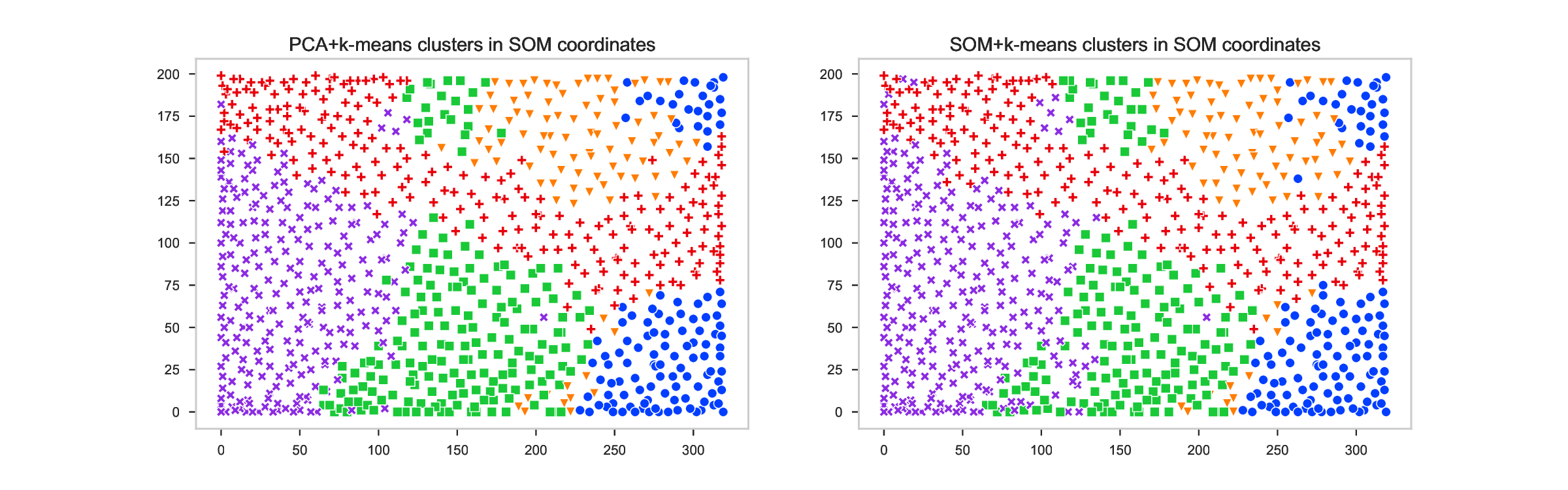
}
\caption{A comparison of the PCA+k-means (left) and SOM (right) clustering. The coordinates are the conditional 2D t-SNE coordinates (top) and the SOM plane (bottom). The points correspond to individual blazars. Different clusters are shown by different colors/symbols.}
\label{fig:comparison}
\end{figure*}

To visualize the clustering results we can use the \mbox{t-dist}\-ri\-buted stochastic neighbor embedding (t-SNE) algorithm \citep{vanDerMaaten2008} or the 2D coordinates derived in the SOM clustering. The (t-SNE) algorithm converts similarities between data points to joint probabilities and tries to minimize the Kullback--Leibler divergence \citep{Kullback51klDivergence} between the joint probabilities in the low-dimensional embedding and the high-dimensional data. In the SOM approach the coordinated are obtained as a result of neural network mapping. The result of our \mbox{PCA\,+\,k-means} 6-dimensional clustering embedded in a 2-dimensional space is shown in the left part of Fig.~\ref{fig:comparison} in the t-SNE (top) and SOM (bottom) coordinates, respectively. The right part of the figure is, accordingly, for the SOM clustering. 

We should note that t-SNE (top panels in Fig.~\ref{fig:comparison}) is a non-linear algorithm focusing on the local similarity of points, and the results also depend on the selection of hyperparameters (mainly, the {\it perplexity}), therefore t-SNE visualization cannot be interpreted as a precise description of object positions in the feature space. For example, the formation of apparently localized groups or some displacement of points belonging to different clusters can be of artificial nature. 
Nevertherless, the figure shows that the results of both the PCA\,+\,k-means (upper left) and SOM (upper right) clusterings are well described by the t-SNE visualization. Some separation of cluster~0 from the general cloud is clearly visible.

The bottom panels are the same PCA\,+\,k-means and SOM clustering on the SOM 2D plane.

Comparing the left and right panels in Fig.~\ref{fig:comparison}, we can conclude that the two methods give similar results and, despite of the absence of localized groups, the boundaries of the clusters are pretty much the same for both methods. 

To compare the results numerically, we used the Rand index \citep{rand1971} calculated in the standard way:
\begin{equation}
R = \frac{2(a+b)}{N(N-1)},
\end{equation}
where $a$ is number of objects that remained in the same cluster after new clustering, $b$ is the number of objects remained in the different clusters, and $N$ is the total number of compared pairs. The Rand index shows what percentage of objects does not change their cluster membership in the two clusterings. For the comparison of the PCA\,+\,k-means and SOM clusterings on the subsample without missing values, we have achieved a Rand index of 0.95.

Additionally, we also tested the approach where \mbox{t-SNE} is used directly to reduce the dimensionality of data, followed by clustering in the 2D or 3D space of t-SNE coordinates. This method gave good results for the sample without missing values; however, for the imputed data and the full sample the results were unsatisfactory.

\subsection{Full sample}

\subsubsection{Missing data imputation}
\label{sec:imputation}

Clustering of the complete BZCAT catalog was carried out according to the same scheme as for the subsample without missing values: dimensionality reduction using PCA followed by k-means clustering. The main difference is the need to fill in the missing values in the model dataset (Fig.~\ref{fig:model_dataset}). We tested several approaches to solve this problem: imputation with the median values, machine learning regression models, namely XGBoost \citep{Chen:2016:XST:2939672.2939785} and scikit-learn \citep{JMLR:v12:pedregosa11a} implementations of Random Forest and Histogram-based Gradient Boosting, and finally the imputation of the missing values using probabilistic PCA \citep{DBLP:journals/neco/TippingB99}, which showed the best results.

The method introduces a latent variable $\bf z$, corresponding to an $M$-dimensional PCA subspace, and probability distributions $p$ such that
\begin{equation}
p({\bf z}) = \mathcal{N}({\bf 0}, {\bf I}),\quad
p({\bf x}) = \mathcal{N}({\bf Wz}+\boldsymbol{\mu}, \sigma^2{\bf I}),
\end{equation}
where $\bf x$ is the observed variable ($D$-dimensional object vector in the feature space) and $\mathcal{N}({\bf 0},{\bf I})$ is the standard normal distribution. The matrix ${\bf W}\in\mathbb{R}^{D\times M}$, vector $\boldsymbol{\mu}$ (equal to zero after scaling), and constant $\sigma^2$ determines the PCA transformation. Probabilistic PCA reduces to classical PCA at $\sigma^2=0$. Such mathematical formalism allows one to determine ${\bf W}$, $\boldsymbol{\mu}$, and $\sigma^2$ via the expectation-maximization (EM) algorithm and impute missing values by sampling from latent $p({\bf z})$. In our case we have used an implementation of pPCA that calculates the imputed values along with ${\bf W}$ and $\sigma^2$ \citep{porta:inria-00321476}, which is considered by the  authors as a more efficient approach.

This final variant (pPCA) for the imputation of missing values was chosen based on the maximum similarity of feature distributions in the obtained clusters for the two samples: (1)~when all missing values had been dropped and (2)~with imputed values. The clustering results of the sample without missing values were used as reference cluster labels.

The similarity of the distributions was estimated by the Kullback–-Leibler divergence \citep{Kullback51klDivergence} calculated in our case as a sum over all feature distributions per each cluster:
\begin{equation}
D_{\rm KL}(P\parallel Q) = \sum_{\mathcal D}\sum_{\mathcal K}\sum_{x\in{\mathcal X}}
    P(x)\log{\frac {P(x)}{Q(x)}},
\end{equation}
where ${\mathcal D}$ are the features in the model dataset, ${\mathcal K}$ are the clusters, 
$P(x)$ and $Q(x)$ are the probability distributions on the sample space ${\mathcal X}$, with $P(x)$ for the full sample and $Q(x)$ for the reference subsample without missing values.

Additionally we evaluated the Rand index between the two clusterings:
$R\simeq90\%$, which is quite good, given the increase in the number of objects by about four times and the fact that cluster boundaries are drawn in a continuous ``cloud'' without clear localization of the groups.

\begin{figure*}
\centering
\includegraphics[width=\textwidth]{
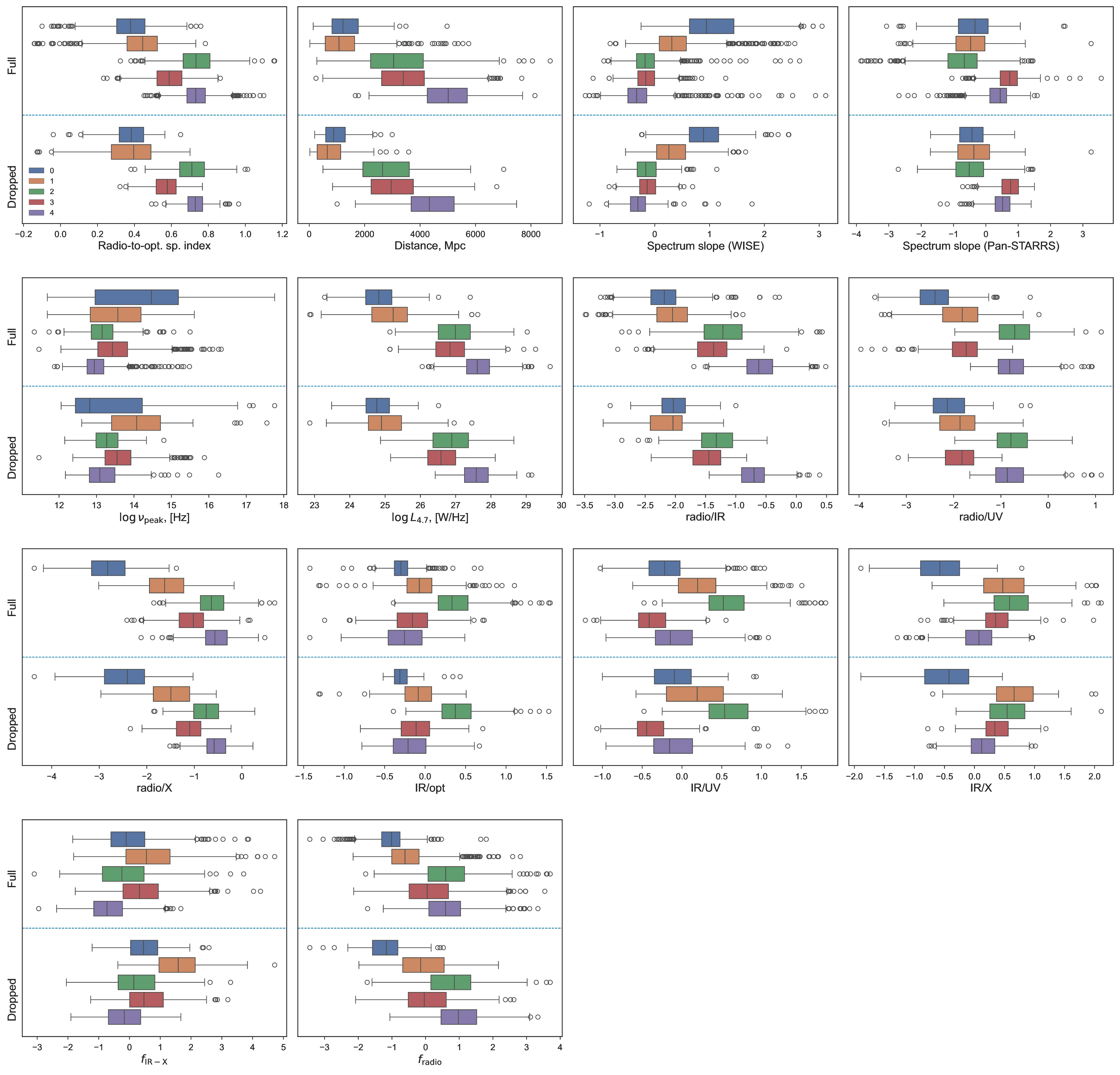
}
\caption{Comparison of the model dataset feature distrubutions for the full Roma-BZCAT sample (upper parts in the panels)
and the subsample without missing values (lower parts in the panels).
The distributions are shown as boxplots. The box is the interquartile range of a parameter distrubution (IQR, 25th to 75th percentile, or Q1 to Q3), the median is shown as a vertical line inside the box, the ``whiskers'' extend to show the rest of the distribution, except for points that are determined as ``outliers,\!'' locating beyond the median~$\pm$~1.5~IQR range, these outliers are shown by dots.
The panels correspond to the features, and the clusters are marked with different colors (see the legend in the first panel). The full sample contains 3527~blazars (we do not consider here the 34 outliers mentioned in Section~\ref{sect:outliers}), while the sample with all the missing values dropped amounts to only 858 objects, a quarter of the full sample. Although we imputed the missing values to perform the clustering for the full sample, these distributions are based on only the real data in both cases.}
\label{fig:distribs}
\end{figure*}

The visual comparison of the model feature distributions in the subsample without missing values and in the full sample is shown in Fig.~\ref{fig:distribs}. 
We can see that the distributions retain their shape well after the substantial increase of the sample. Notice also that the observed discrepancies for some parameters (e.g., a substantial difference in the median $\log\nu_{\rm peak}$ values for cluster~0) should not be treated as errors, as the membership of new objects in the clusters is defined based on the entire number of features, and some changes in the shape of the distributions are expected after increasing the sample by several times. For instance, the strongest observed difference in the $\log\nu_{\rm peak}$ distributions is in good agreement with the lowest importance of this feature for the clustering (see Table~\ref{tab:importance}).

We also performed the clustering of the full dataset by the SOM method, which showed a Rand index of $\sim0.92$ with respect to the PCA+k-means results. Thus, the two clustering methods showed a 92\%\ similarity for the complete dataset along with even better concordance for the smaller dataset without missing values (95\%).
The similar results also prove that there are no nonlinearities in the data distribution, which could not be taken into account by the PCA+k-means method. Therefore, PCA+k-means has been used for further analysis, as a more straightforward approach.

\begin{figure}
\centering
\includegraphics[width=\columnwidth]{
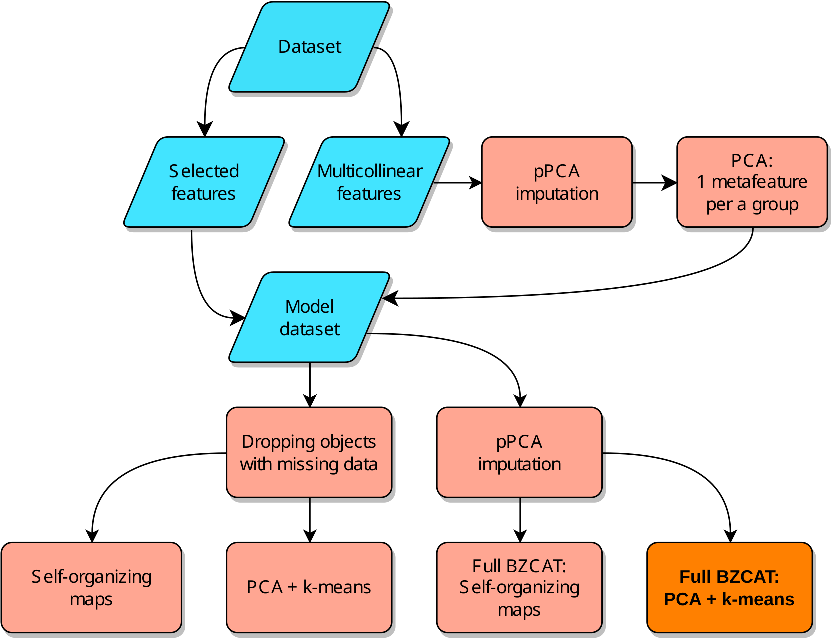
}
\caption{The flowchart of the performed clustering stages.}
\label{fig:flowchart}
\end{figure}

The totality of the results obtained allows us to perform the cluster analysis not only for the subsample without missing values but for all blazars in the Roma-BZCAT catalog. The flowchart of the clustering stages is shown in Fig.~\ref{fig:flowchart}, the final result for the full sample is highlighted in the lower right corner of the figure.

\subsubsection{Classification of outliers}
\label{sect:outliers_class}

To include all the Roma-BZCAT blazars in the clustering results, the membership of the 34~outliers filtered out at the data cleansing stage (Section ~\ref{sect:outliers}) must be determined.

We used the k-nearest neighbors classifier from the {\tt scikit-learn} library \citep{JMLR:v12:pedregosa11a} to complete the task. The dataset with the obtained cluster labels, which acted as a target variable for training the classifier, was divided into the training and test samples in a ratio of 0.1. The hyperparameters were optimized by 5-fold cross-validation on the training sample using a grid search: the number of nearest neighbors in the $[5,100]$ range without weighting and with distance-based weights. The quality metric was the F1-score. The final hyperparameters, 70 nearest neighbors with weights inversely proportional to the distance, gave an F1-score of $\simeq0.94$ on the test sample: the harmonic mean of the precision and recall for the trained classifier reaches 94\%.

For the trained classifier to work correctly, all the stages of data preprocessing for the outliers must be performed in the same way as for the 
training dataset, but here we could not use probabilistic PCA to replace missing values since the pPCA implementation we used computed them along with PCA transformation parameters, which must be fixed during the inference.
For that reason, here we used the following approach:
\begin{itemize}
\item instead of the first pPCA step (for the multicollinear flux densities transformed to metafeatures), we took the mean value over the corresponding flux densities for each object;
\item for the second pPCA step, the missing values in the model dataset were imputed  as the mean values over a column (which is actually zero after the scaling);
\item other transformations (scaling, traditional PCA, etc.) corresponded to the main clustering model.
\end{itemize}

\pagebreak

\subsection{Robustness of the clustering to dataset incompleteness 
and feature selection}
\label{sec:robustness}

Two conditions that can influence the obtained results are the incompleteness of the Roma-BZCAT sample and the features we selected to perform the clustering. The former can change the boundaries of the clusters after taking a sufficient enough amount of new blazars, and the latter could form a new feature space with additional information about the objects. Particularly, in our clustering dataset we did not take into account the characteristics connected with the gamma-ray emission in the gamma-ray ranges, but blazars emit a large amount of their radiation in gamma-rays, which means that the gamma-ray band should carry important information about the sources.

The gamma-ray measurements, though, are too scarce to be used in the clustering of the whole Roma-BZCAT blazars: the Fermi~LAT gamma-ray flux in the catalog is given for only 28\%\ of the sources. The new Fermi~LAT measurements \citep{2022ApJS..263...24A} do not fundamentally change the situation: we evaluated that now 44\%\ of the objects would have gamma-ray fluxes, which is still insufficient (over 60\%--70\%\ of available data are needed.) 

Nevertheless, using the present gamma-ray data, we can still evaluate the degree to which the use of gamma-ray measurements is able to change the obtained clustering results as well as evaluate the influence of  dataset incompleteness. To this end, we took only the objects with available gamma-ray measurements and calculated for this subsample additional gamma-ray features analogously to our previously described data preparation. The added features are gamma-ray flux, luminosity, and hardness ratios relative to other spectral ranges, a total of 7 new features to complement the 14 already available. After dropping the missing values for all the 21 features, we end up with a small dataset of 396 sources. Thus, we, firstly, shorten the list of objects to as few as 11\%\ of the complete sample and, secondly, add 50\%\ new features with sufficiently different information concerning the gamma-ray range. To separate the influence of the two effects, we (1)~compared the results of the clustering performed with 14 original features on the small dataset and on the complete sample; (2)~compared the results of the clustering performed with 14 original features on the small dataset and the results obtained on the same dataset with 21~features. The Rand indices for these two comparisons are 0.85 and 0.80, respectively; i.e., a $\sim\!90\%$ incompleteness of the sample could change the result at about $1-0.85=15\%$, while the addition of 50\%\ new features preserve it at a level of about 80\%.

From the above evaluations, we can state that clustering labels for the sources should stay the same within 80\%\ of the current clustering results if new sufficient data on gamma-ray fluxes are going to be available in the future. We also evaluated the importance of the features if taken with the gamma-ray range, the result is presented in Table~\ref{tab:importance_gamma}. As one can see, the gamma features occupy the upper rows of the table, thus proving that the gamma range is important for blazar classification and further more abundant measurements would lead to better, more accurate, results.

\begin{table}
\centering
\caption{
Importance of the features when taking into account the gamma-ray range
}
\label{tab:importance_gamma}
\small
\begin{tabular}{lc}
\hline
Feature & Contribution, \% \\
\hline
Radio/gamma                 & 7.7 \\
$f_{\gamma}$                & 6.5 \\
Radio/IR                    & 6.2 \\
UV/gamma                    & 6.1 \\
Opt/gamma                   & 5.6 \\
X/gamma                     & 5.4 \\
$\log L_{5}$                & 5.2 \\
Radio-to-opt. sp. index     & 5.0 \\
Radio/X                     & 5.0 \\
$\log L_{\gamma}$            & 4.7 \\
Spectrum slope (WISE)       & 4.6 \\
Distance                    & 4.6 \\
Radio/UV                    & 4.5 \\
IR/gamma                    & 4.4 \\
IR/UV                       & 4.3 \\
$f_{\rm radio}$             & 4.1 \\
Spectrum slope (Pan-STARRS) & 3.7 \\
IR/opt                      & 3.6 \\
IR/X                        & 3.3 \\
$f_{\rm IR-X}$              & 3.1 \\
$\log \nu_{\rm peak}$       & 2.3 \\
\hline
Total                       & 100 \\
\hline
\end{tabular}
\end{table}

\section{Properties of the clusters}

\subsection{Comparison with the known classes}
\label{sec:compare_classes}

First of all, it is interesting to compare our clusters with the known types of blazars. Here we make such a comparison for the Roma-BZCAT blazar types, for the high synchrotron peaked (HSP) blazars from the 3HSP catalog, and for the blazars detected in the TeV energy range from the TeVCat catalog.

In the Roma-BZCAT catalog, blazars are divided into the following subtypes.
\begin{description}
\item[BZB] BL\,Lac objects and BL\,Lac candidates, which are AGNs with a featureless optical spectrum or having only absorption lines of the host galaxy origin and weak narrow emission lines;
\item[BZG] sources usually reported in the literature as BL\,Lac objects but having SEDs with significant dominance of host galaxy emission;
\item[BZQ] Flat Spectrum Radio Quasars (FSRQs) with the optical spectrum showing broad emission lines and dominant blazar characteristics;
\item[BZU] blazars of an uncertain type, a small number of sources having peculiar characteristics but also exhibiting blazar activity: the occasional presence/absence of broad emission lines or other features, transition between a radio galaxy and a BL\,Lac, galaxies hosting a low luminosity blazar nucleus, etc. 
\end{description}

\begin{figure*}
\includegraphics[width=\textwidth]{
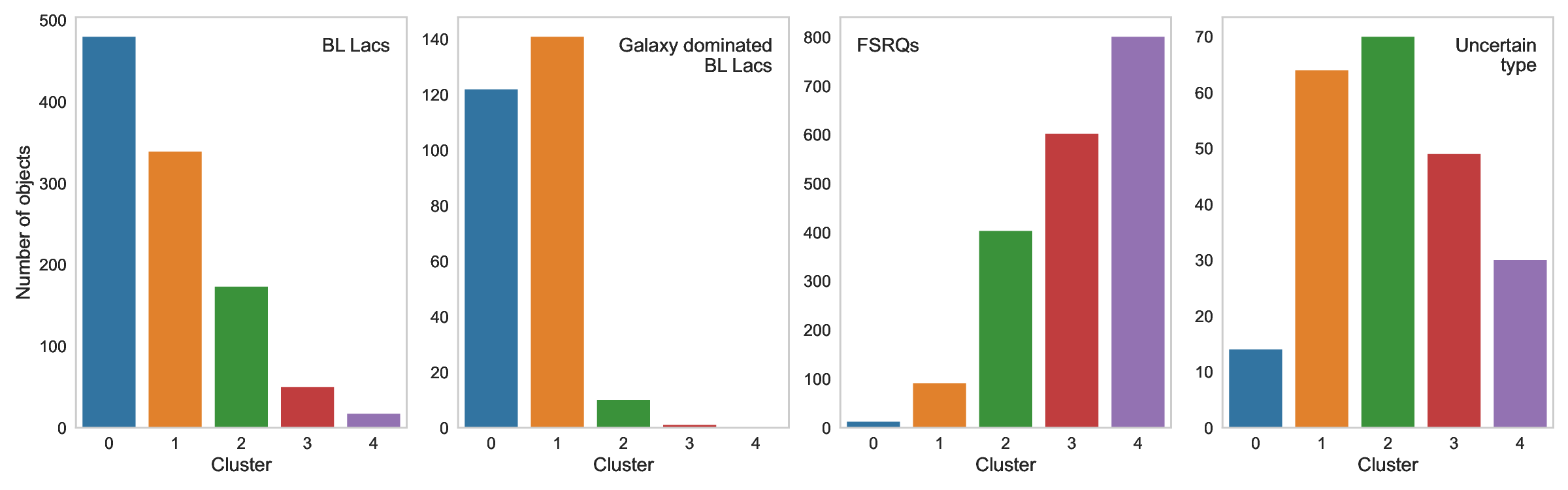
}
\caption{Cross identification with the Roma-BZCAT classes. Each panel corresponds to a certain blazar type and shows the number of the blazars of this type within the clusters.}
\label{fig:crossident}
\end{figure*}

In Table~\ref{tab:cross} and Fig.~\ref{fig:crossident} we compare the population of the obtained clusters with the subtypes of blazars in Roma-BZCAT. The vast majority of BL\,Lacs and BZGs fall into clusters 0 and 1. Clusters 3 and 4, on the contrary, are dominated by FSRQs. Cluster 2 is a mixture of BL\, Lacs and FSRQs. Blazars of an uncertain type avoid cluster 0 and are less present in clusters 3 and 4. It is noteworthy that the largest number of them are in cluster 2, a mixture of BL\,Lacs and FSRQs,
although a comparable number is found in cluster~1.

\begin{table}
\caption{Cross identification with the Roma-BZCAT classes. BZGs are the galaxy-dominated BL\,Lacs}
\label{tab:cross}
\small
\begin{tabular}{rrrrrr|r}
\hline
        & \multicolumn{6}{c}{BZCAT classes} \\
\cline{2-7}
Clust. & BL\,Lac &  BZG  & BL\,Lac & FSRQ & Uncert. & Total\\
        &         &       & cand.   &    {}  &     {}    & \\ 
 \hline
0 & 480 & 122 & 55 &  12 & 14 & 683 \\
1 & 339 & 141 & 14 &  91 & 64 & 649 \\
2 & 173 &  10 & 11 & 403 & 70 & 667 \\
3 &  50 &   1 &  8 & 602 & 49 & 710 \\
4 &  17 &   0 &  4 & 801 & 30 & 852 \\
\hline
Total & 1059 & 274 & 92 & 1909 & 227 & 3561\\
\hline
\end{tabular}
\end{table}

Judging by the quality metrics obtained earlier, we assign blazars to a particular cluster with an accuracy of about 90\%, therefore a small number of blazars of the ``opposite'' types in individual clusters, with the exception of cluster 2, can be considered expected. Taking into consideration the correlated continuous decrease/increase of the number of BL\,Lacs and FSRQs among the clusters, it also could be a real effect to some degree. In total, we can state that the clustering results largely correlate with the classification of blazars in the Roma-BZCAT catalog.
At the same time, our clustering additionally distinguishes between two subclasses of BL\, Lacs (clusters 0, 1) and two subclasses of FSRQs (clusters 3, 4). There is also no division into BL\, Lacs and galaxy-dominated BL\,Lacs, although the almost complete absence of the latter in the ``mixed'' cluster~2 could be noted.

Blazars are classified into a separate type of AGNs since they have a distinct orientation of the jet, pointing toward the observer at a small angle. As well as other AGNs, they have similar structure (e.g., a supermassive black hole, an accretion disk, a jet, etc.) and similar processes (the collimation of the jet, acceleration of electrons in a magnetic field, accretion of matter onto the central object) 
but occurring under different physical conditions, which causes their division into different subclasses according to the observed parameters. 
Thus, FSRQs have strong emission lines and higher luminosity compared to BL\,Lacs in almost all frequency ranges, which is probably related to the more abundant fueling matter and, consequently, different accretion modes.
The fact that different blazar types are not isolated in our clusters but demonstrate a continuous per-cluster distribution validates the commonly accepted uniformity of blazars' nature. Notice also that although we intentionally avoided any predetermined categorical separation such as the presence or absence of emission lines, the clustering correlates with the BL\,Lac/FSRQ classification, thus proving that this difference in physical conditions can be obtained from other characteristics.

\begin{figure}
\centering
\includegraphics[width=\columnwidth]{
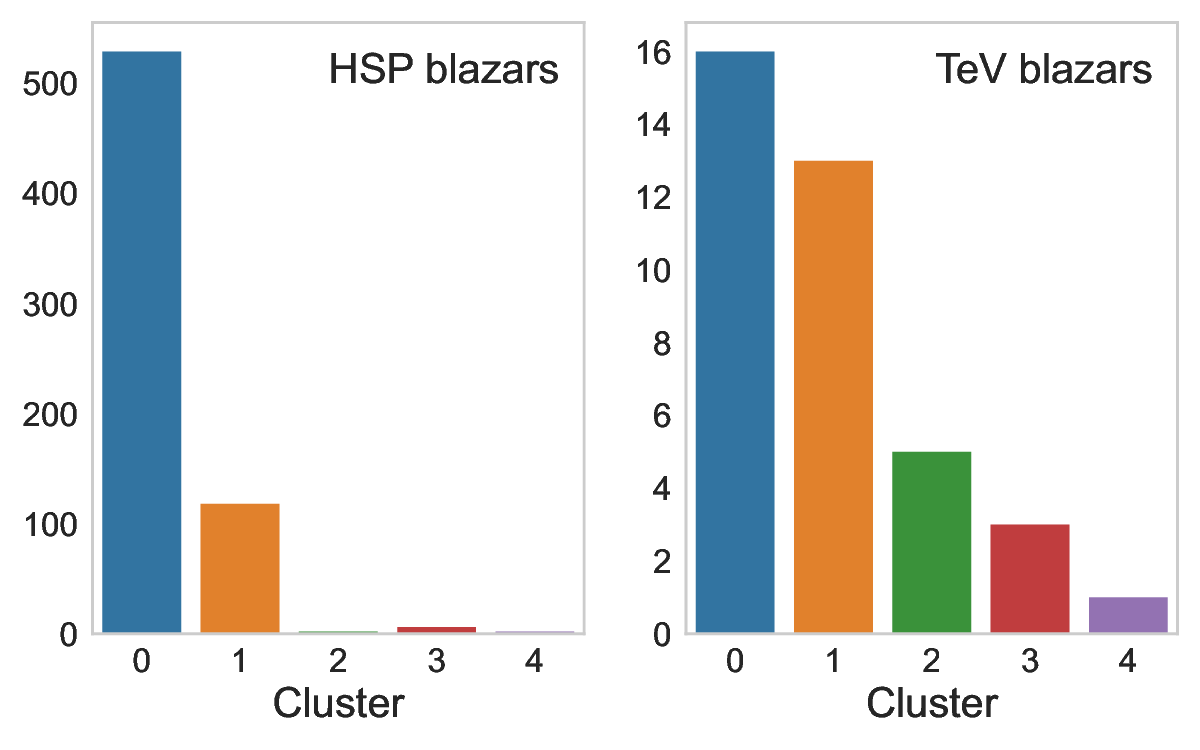
}
\caption{Cross identification with the high synchrotron peaked blazars from the 3HSP catalog and the blazars detected in the TeV energy range from the TeVCat catalog. The panels are organized analogously to Fig.~\ref{fig:crossident}}
\label{fig:hsp_tev_barchart}
\end{figure}

\begin{figure*}
\centering
\includegraphics[width=\columnwidth]{
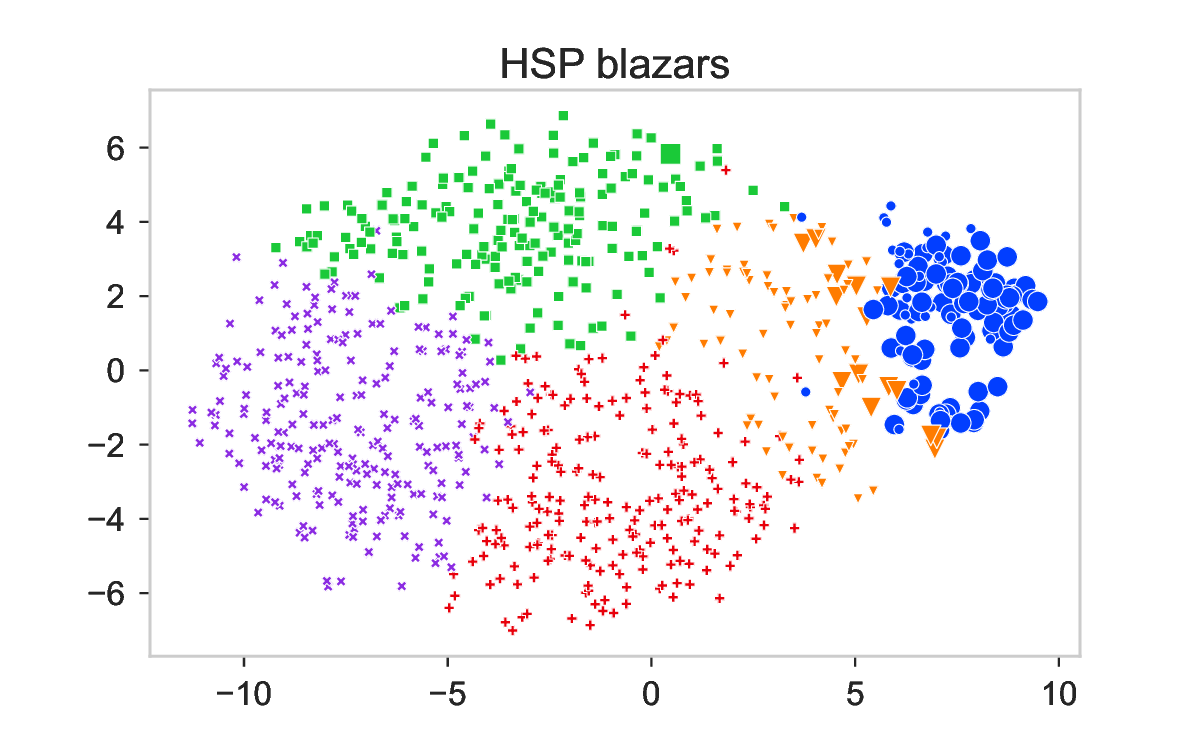
}
\includegraphics[width=\columnwidth]{
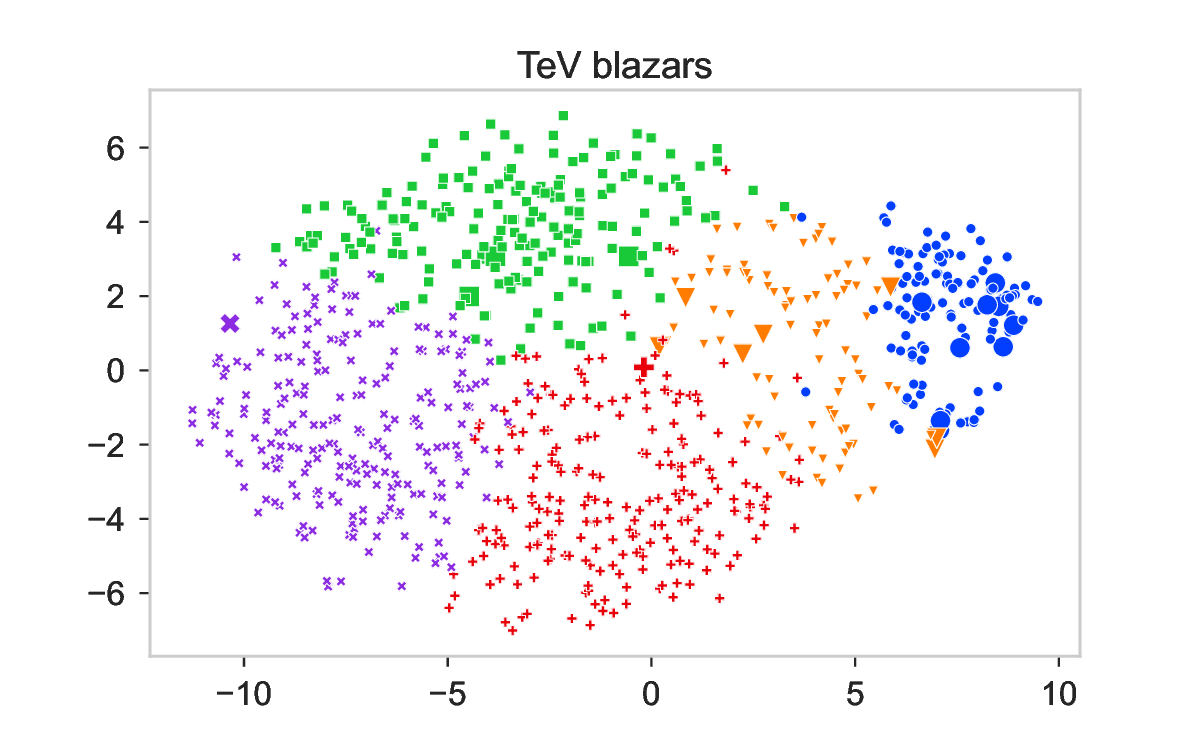
}
\caption{High synchrotron peaked (HSP) and TeV blazars on the t-SNE cluster map. The points correspond to individual blazars. Different clusters are shown by different colors/symbols. The bigger symbols correspond to the HSP (left panel) and Tev (right panel) blazars.}
\label{fig:hsp_tev_map}
\end{figure*}

In Figs.~\ref{fig:hsp_tev_barchart} and~\ref{fig:hsp_tev_map} we demonstrate how the high synchrotron peaked (HSP) blazars from the 3HSP catalog \citep{2019A&A...632A..77C} and the blazars detected in the TeV energy range from the TeVCat catalog \citep{2008ICRC....3.1341W} are distributed within our clusters. Figure~\ref{fig:hsp_tev_map} clearly show that almost all HSP blazars are members of cluster~0, and the rest of them, located in cluster~1, lay closer to the boundary of the two clusters. The TeV blazars are not that concentrated to a particular cluster, but have a tendency to be more abundant in BL\,Lac-populated clusters~0--1 than in FSRQ-populated clusters~3--4. The overall number of TeV blazars is small, and there are only few of them found in the latter clusters, so their presence in the FSRQ-populated clusters is questionable and may be caused by clustering inaccuracy; at the same time the descent of the number of TeV blazars from cluster~0 to cluster~4 in Fig.~\ref{fig:hsp_tev_barchart} looks pretty smooth.

\subsection{Description of the clusters}

\begin{figure*}
\includegraphics[width=\textwidth]{
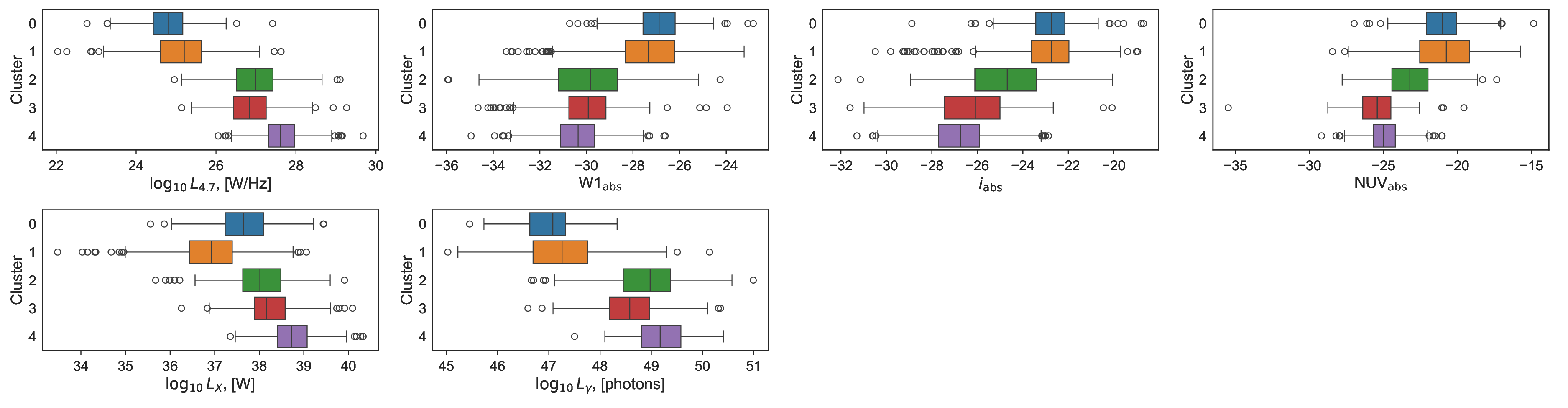
}
\caption{The distributions of luminosities or absolute magnitudes in different ranges of the electromagnetic spectrum. The boxplots are constructed analogously to Fig.~\ref{fig:distribs} Each panel corresponds to a particular characteristic, the distributions within the clusters are shown by different colors along the $y$-axes.
Warning: the radio luminosities are corrected for different redshifts using radio spectral indices, the other values are not.}
\label{fig:luminosities}
\end{figure*}

\begin{figure*}
\includegraphics[width=\textwidth]{
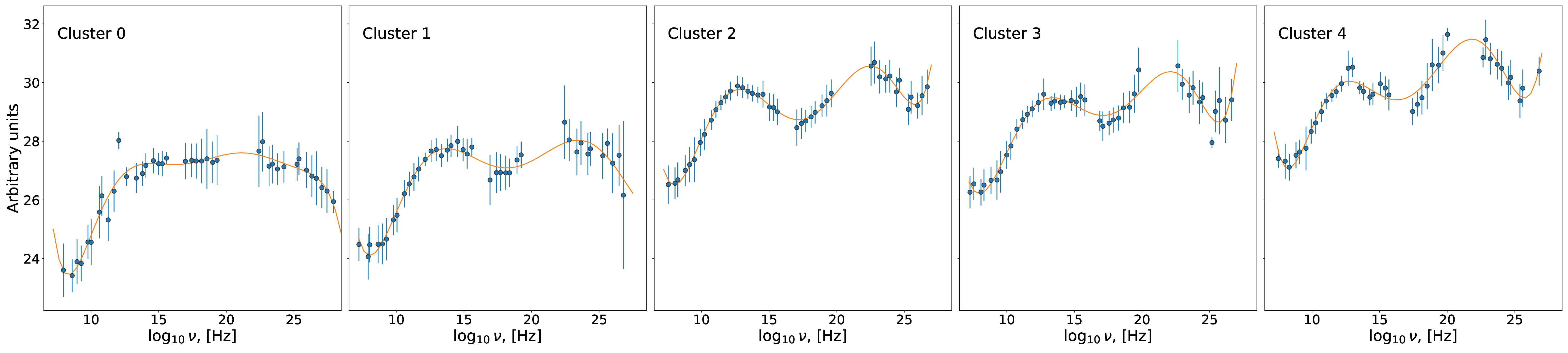
}
\caption{Average SEDs for the clusters in the rest frame. The frequency range is from radio waves (lower abscissa values) to gamma-ray emission.
The $y$-axis is scaled in such a way that the median luminosity $\log_{10}L_{5}$ corresponds to the 5~GHz flux density of the average SED. The SEDs are fitted using 7th-degree polynomials (the orange lines).}
\label{fig:cl_seds}
\end{figure*}

To investigate the properties of the selected groups, we analyzed the differences in the distributions of blazar characteristics in the obtained clusters. Partially, these differences are demonstrated in Fig.~\ref{fig:distribs}. Some statistics of the distributions are given in Table~\ref{tab:cluster_stats}.
Additionally, in Fig.~\ref{fig:luminosities} the luminosities or absolute magnitudes for different spectral ranges are given. The values should be used with caution: while the radio luminosities are corrected for different redshifts using radio spectral indices, the other values are not. Therefore, for instance, for high-redshift blazars the optical absolute magnitude in the $i$ filter actually corresponds to the UV range in the source's frame of reference. The effect is strong for the IR and optical ranges, but is less significant for X-rays and gamma-ray luminosities as they are measured in broad bands and anyway stay withing the corresponding electromagnetic ranges at any redshifts, drifting though to higher frequencies. A better understanding of the ratios between flux densities in different ranges of the electromagnetic spectrum without the $z$-dependent bias may be obtained from Figure~\ref{fig:cl_seds}, which shows the rest-frame average SEDs for the resulting clusters.

\begin{table*}
\caption{Cluster characteristics for the whole Roma-BZCAT catalog}
\label{tab:cluster_stats}
\small
\begin{tabular}{lc|ccccc}
\hline
 & & \multicolumn{5}{c}{Medians (Min--max values)}  \\
\cline{3-7}
 &  & Rad.-opt. & $d$, Mpc & $\log_{10}\nu_{\rm peak}$, [Hz] & $\log_{10}L_{5}$, [W/Hz] & $\log_{10}(\nu F_{1.4}/\nu F_{\rm W2})$ \\
 
 Cluster & $N$ & sp.\ index  &  & (synchrotron peak) & (radio luminosity) & (radio/IR)  \\ 
\hline  
 0 (BL\,Lacs) & 683 & 0.4 ($-0.4$, $+0.8$) & 1200 (120, 4970)  &
  14.5 (11.7, 18.4) & 24.8 (22.8, 27.4) & $-2.2$ ($-3.2$, $-0.3$)\\
 
 1 (BL\,Lacs) & 649 & 0.4 ($-0.3$, $+0.8$) & 1050 (10, 5750)   &
  13.6 (11.7, 15.6) & 25.2 (22.0, 27.6) & $-2.0$ ($-3.8$, $-0.9$)\\
 
 2 (mix)      & 667 & 0.7 ($+0.3$, $+1.2$) & 3050 (270, 8690)  &
  13.1 (11.3, 15.5) & 27.0 (25.0, 29.1) & $-1.2$ ($-3.6$, $+0.4$)\\
 
 3 (FSRQs)    & 710 & 0.6 ($+0.2$, $+0.9$) & 3390 (80, 7670)   & 
  13.4 (11.5, 16.3) & 26.8 (25.1, 29.3) & $-1.4$ ($-3.0$, $-0.2$)\\
 
 4 (FSRQs)    & 852 & 0.7 ($+0.5$, $+1.2$) & 5010 (1660, 8140) &
  12.9 (10.4, 15.5) & 27.6 (26.1, 29.7) & $-0.6$ ($-1.7$, $+1.5$)\\
\hline
 &  & $\log_{10}(\nu F_{1.4}/\nu F_{\rm NUV})$  & $\log_{10}(\nu F_{1.4}/\nu F_{\rm X})$ & $\log_{10}(\nu F_{1.4}/\nu F_{\gamma})$ & $\log_{10}(\nu F_{\rm W2}/\nu F_{i})$ & $\log_{10}(\nu F_{\rm W2}/\nu F_{\rm NUV})$ \\
 
 &  & (radio/UV) & (radio/X-rays) & (radio/$\gamma$-rays) & (IR/optical) & (IR/UV) \\

 0 (BL\,Lacs) & 683 & $-2.4$ ($-3.7$, $-0.4$) & $-2.8$ ($-4.4$, $-1.4$)
  & $-4.1$ ($-5.2$, $-3.1$) & $-0.3$ ($-1.4$, $+0.7$) & $-0.2$ ($-1.0$, $+1.0$) \\
 
 1 (BL\,Lacs) & 649 & $-1.8$ ($-3.5$, $-0.2$) & $-1.6$ ($-3.4$, $-0.2$)
  & $-3.7$ ($-5.0$, $-2.2$) & $-0.1$ ($-1.3$, $+1.1$) & $+0.2$ ($-0.6$, $+1.5$) \\
 
 2 (mix)      & 667 & $-0.7$ ($-2.0$, $+1.1$) & $-0.6$ ($-1.9$, $+1.5$)
  & $-3.2$ ($-4.5$, $-2.1$) & $+0.3$ ($-0.4$, $+1.8$) & $+0.5$ ($-0.5$, $+2.4$) \\
 
 3 (FSRQs)    & 710 & $-1.7$ ($-4.0$, $-0.8$) & $-1.0$ ($-2.4$, $+0.2$)
  &  $-3.3$ ($-4.4$, $-1.8$) & $-0.2$ ($-1.2$, $+0.7$) & $-0.4$ ($-1.8$, $+0.5$) \\
 
 4 (FSRQs)    & 852 & $-0.8$ ($-1.6$, $+0.9$) & $-0.6$ ($-2.1$, $+1.0$)
  & $-3.0$ ($-4.5$, $-1.8$) & $-0.3$ ($-1.4$, $+0.7$) & $-0.1$ ($-1.0$, $+1.1$) \\
\hline
 & & $\log_{10}(\nu F_{\rm W2}/\nu F_{\rm X})$ & $\log_{10}(\nu F_{\rm W2}/\nu F_{\gamma})$ & $\log_{10}(\nu F_{i}/\nu F_{\gamma})$ & $\log_{10}L_{X}$, [W] & $\log_{10}L_{\gamma}$, [ph/s] \\
 
 & & (IR/X) & (IR/$\gamma$-rays) & (optical/$\gamma$-rays) & (X-ray lumin.) & ($\gamma$-ray lumin.)\\

 0 (BL\,Lacs) & 683 & $-0.6$ ($-1.9$, $+0.8$) & $-1.8$ ($-3.1$, $-1.1$) 
 & $-1.4$ ($-2.9$, $-0.7$) & 37.6 (35.6, 39.4) &  47.1 (45.5, 48.3) \\
 
 1 (BL\,Lacs) & 649 & $+0.5$ ($-0.7$, $+2.7$) & $-1.6$ ($-2.7$, $-0.1$) & $-1.5$ ($-2.9$, $+0.2$) & 36.9 (33.5, 39.1) & 47.3 (45.0, 50.1) \\
 
 2 (mix)      & 667 & $+0.6$ ($-0.5$, $+2.5$) & $-1.8$ ($-3.4$, $-0.8$) & $-2.2$ ($-4.0$, $-0.6$) & 38.0 (35.7, 39.9) & 49.0 (46.7, 51.0) \\
 
 3 (FSRQs)    & 710 & $+0.3$ ($-0.9$, $+2.0$) & $-1.8$ ($-3.0$, $-0.7$) 
 & $-1.9$ ($-3.1$, $-0.6$)  & 38.2 (36.3, 40.1) & 48.6 (46.6, 50.4) \\
 
 4 (FSRQs)    & 852 & $+0.1$ ($-1.3$, $+1.0$) & $-2.2$ ($-3.6$, $-1.4$) & $-2.1$ ($-3.7$, $-1.3$)  & 38.7 (37.4, 40.3) & 49.2 (47.5, 50.4)\\
\hline
\end{tabular}
\end{table*}

To construct the average SEDs, we normalized the SEDs of individual blazars by the measured synchrotron peak flux density (Section~\ref{sec:character_calc}, Fig.~\ref{fig:sed_poly}) and averaged the flux densities in 50~bins. The error bars in Fig.~\ref{fig:cl_seds} correspond to the standard deviation within the bins. 
To demonstrate the difference in luminosities, the normalized spectra are additionally adjusted to the radio luminosity at a frequency of 5~GHz
($\log_{10}L_{5}$). The SEDs are recalculated to the rest frame.

\begin{figure*}
\centering
\includegraphics[width=0.8\textwidth]{
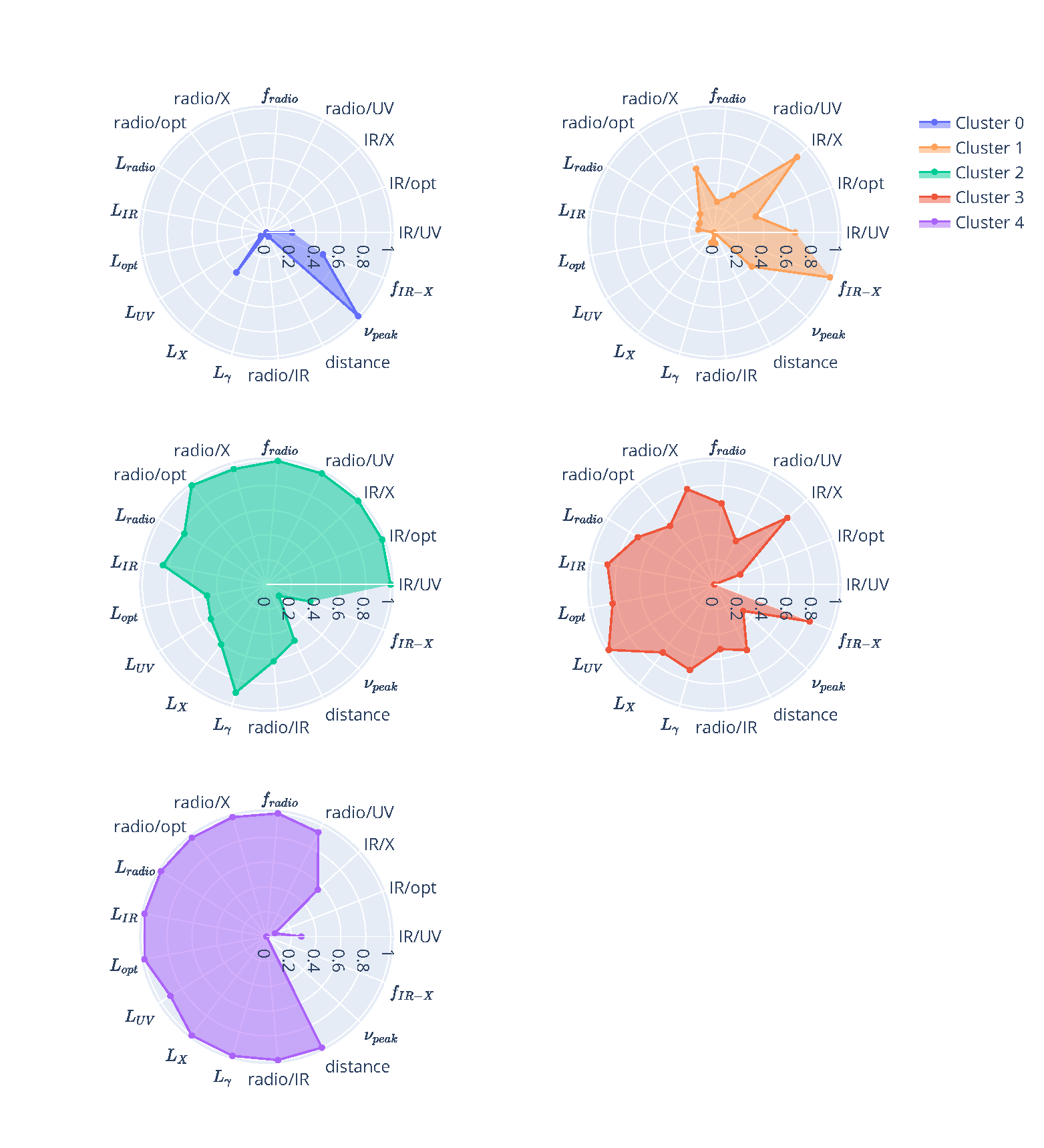
}
\caption{Polar diagrams for scaled medians. Clusters~0--4 are shown from left to right and from top to bottom, also with different colors. The outer edges of the circles correspond to the highest median values, while the centers are the minimum median values. Each ``azimuth'' corresponds to a particular characteristic.}
\label{fig:polar_diag}
\end{figure*}

To better visualise the difference between cluster statistics demostrated in Figs.~\ref{fig:distribs},~\ref{fig:luminosities} and Table~\ref{tab:cluster_stats}, we also constructed polar diagrams shown in Fig.~\ref{fig:polar_diag}. The figure reflects the difference between median values of various characteristics: the polar diagrams are scaled in such a way that the maximum observed medians over all clusters correspond to values of 1 (the outer edges of the circles), while the minimum medians correspond to zero values (the centers of  the circles). Each ``azimuth'' corresponds to a particular characteristic. 

Before describing the average SEDs, we first should mention the following.
The SED of a blazar has typically a shape of two humps and constitutes a complex mix of emission from different parts of an active galactic nucleus. The first hump, extending from radio waves to X-rays is believed to be formed by the synchrotron radiation in the jet. Part of the photons emitted in this process may experience synchrotron self-Compton scattering (e.g., \citealt{1996ApJ...461..657B}) contributing to the second, gamma-ray, hump. Photons from other AGN components such as the accretion disk, dust torus, and broad emission line clouds also contribute to the gamma-ray hump via inverse Compton scattering. The location of this gamma-ray emission is still a subject of research (e.g., \citealt{2016JPhCS.718e2032R}). Additionally, the accretion disk emits its own thermal radiation peaked in the optical--UV ranges, while the dust torus adds to the IR emission. The corona of the accretion disk can also scatter photons up to the X-ray energies. At last, the SED may be affected by the host galaxy.

All this complexity results in that the detailed description can only be made for a particular SED via complex modelling and/or analysis of its variability time series for different ranges of the electromagnetic spectrum. Nevertheless, in the case of blazars, we have an advantage that 
their jets are inclined with respect to the line of sight at a small angle, therefore we can expect that the differences in the average SEDs are caused to a greater degree not by the geometric effects but by the different physical conditions in the AGN, whatever these conditions are.

\subsubsection{Clusters 0 and 1: BL\,Lac subclasses}

These two clusters consist of BL\,Lacs and galaxy-dominated BL\,Lacs located at relatively small distances (up to 3~Gpc, $z\lesssim0.9$). The percentage of FSRQs is only 2\%\ and 14\%\ respectively, see Table~\ref{tab:cross}.
Cluster blazars are distinguished by relatively reduced luminosity in the entire range of the electromagnetic spectrum (Table~\ref{tab:cluster_stats}, Figs.~\ref{fig:luminosities} and \ref{fig:cl_seds}). Also, they have significantly reduced radio hardness parameters (Table~\ref{tab:cluster_stats}, Fig.~\ref{fig:distribs}). As well, the objects have lower gamma-ray luminosities (the same table and figures).

Cluster 0 has the most prominent characteristics. From Fig.~\ref{fig:cl_seds} we can see that both the synchrotron and gamma-ray humps are factually not visible in the  average SED. 
By reviewing some of the individual SEDs in the cluster, we have found that actually they have a standard shape of two humps. Therefore, this effect in the average SED is caused by the broad distribution of synchrotron peaks, seen in Fig.~\ref{fig:distribs}.

The average SED of cluster~1 has the classical shape of two humps, nevertheless with both humps not that prominent. The difference in shape with cluster~0 is likely caused by a more compact distribution of synchrotron peak frequencies (Fig.~\ref{fig:distribs}).

This broad variation in the synchrotron peak frequency, especially noticeable in cluster~0 (see Fig.~\ref{fig:distribs}), in the entire range of \mbox{$\log\nu\simeq11.7$--$18.4$}, is a characteristic peculiarity of the clusters, in other clusters the distributions are more compact, and high frequencies of the synchrotron peak are practically not found. One can notice from Fig.~\ref{fig:distribs} that almost all high synchrotron peaked blazars (HSPs) should belong to cluster 0. Earlier in Section~\ref{sec:compare_classes} we considered the distribution of sources from the 3HSP catalogue \citep{2019A&A...632A..77C} across our clusters and confirmed that 529 of 657 3HSP blazars presented in Roma-BZCAT belong to cluster~0, and 118 to cluster~1, while only 10 are found in clusters 2, 3, and 4 (see Fig.~\ref{fig:hsp_tev_barchart}). Notice also that the 3HSP blazars from cluster~1 are located close to the boundary of clusters~0 and~1.

Notice that within $z\lesssim0.9$, where the blazars of clusters~0 and 1 are located, the actual frequency emitted by a blazar becomes higher for large $z$, but all the frequencies used in the clustering feature space remain within their electromagnetic bands: 3.35~{\textmu}m~$\Rightarrow$~1.76~{\textmu}m~(IR), 7520~\AA~$\Rightarrow$~3960~\AA~(optics), 2316~\AA~$\Rightarrow$~1220~\AA~(UV) for the most distant objects. The radio, UV, X-ray, and gamma-ray frequencies remain within their bands for all $z$ considered in this paper. For the average SEDs, the flux densities were transformed to the rest frame beforehand.

\subsubsection{Cluster 2: BL\,Lac--FSRQ mix}

Cluster~2 is represented by a mixture of BL\,Lac (29\% with BL\,Lac candidates) and FSRQ (60\%) blazars, the remaining 11\% of the objects are of an uncertain type (Table~\ref{tab:cross}). It practically does not contain galaxy-dominated BL\,Lacs, a small number of them (10 objects) can be attributed to clustering errors.

\begin{figure}
\centering
\includegraphics[width=\columnwidth]{
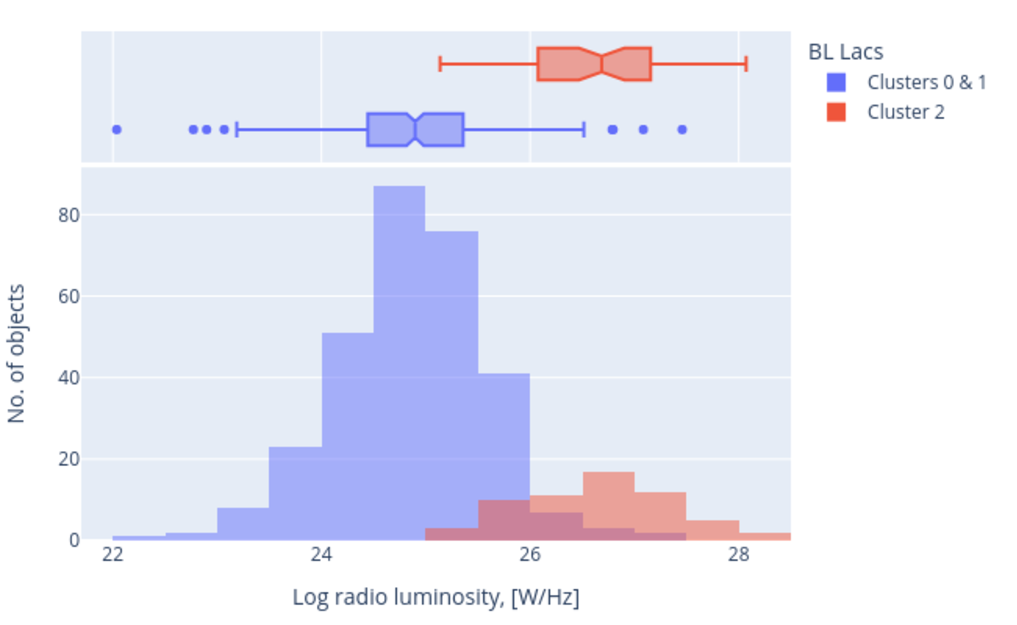
}
\caption{Difference in radio luminosity distributions of the low-luminosity BL\,Lacs in clusters~0--1 (blue color) and the high-luminosity BL\, Lacs in cluster~2 (red color). The boxplots for the distributions are shown at the top.}
\label{fig:lum_dif}
\end{figure}

In contrast to clusters 0 and 1, in cluster 2 we observe high radio, X-ray, and gamma-ray luminosities  as well as bright absolute IR magnitudes comparable to other FSRQs (Figs.~\ref{fig:luminosities} and~\ref{fig:cl_seds}). The absolute magnitudes in the optical and UV ranges are somewhat weakened compared to other FSRQs. 
In this connection, it is of interest to evaluate if the BL Lacs in cluster~2 are anyhow different from those in clusters~0 and 1 or the statistically higher luminosities are only related to the presence of FSRQs. In Fig.~\ref{fig:lum_dif} we compare the radio luminosity distributions for the BL\,Lacs, and the observed difference clearly testifies that BL Lacs from cluster~2 are a special BL\,Lac subclass with high luminosity.

The average SED of the cluster in Fig.~\ref{fig:cl_seds} shows the most smooth shape of classical two humps. The distributions and statistics in Figs.~\ref{fig:distribs},~\ref{fig:luminosities} and Table~\ref{tab:cluster_stats} naturally follows from this shape. 

Within the observed redshifts for cluster blazars, \mbox{$z\sim 0.05$--$2.5$}, the IR radiation, when converted to the rest frame of the source, remains generally within the IR range, approaching optical wavelengths for the most distant objects: 3.35~{\textmu}m~$\Rightarrow$~0.96~{\textmu}m; optical radiation moves into the UV range with growing distance: 7520~\AA~$\Rightarrow$~2150~\AA; UV radiation becomes harder: 2316~\AA~$\Rightarrow$~660~\AA. Again, the average SEDs in Fig.~\ref{fig:cl_seds} are calculated in the rest frame and do not suffer from differing redshifts.

\subsubsection{Clusters 3 and 4: FSRQ subclasses}

Clusters 3 and 4 are populated by FSRQs: 85\%\ and 94\%, respectively, or 91\%\ and 97\%\ if we exclude blazars of an uncertain type (see Table~\ref{tab:cross}). Blazars from these clusters have high luminosities in the entire frequency range. The main difference between clusters~3, 4, and the upper described cluster 2, as it can be seen from the average spectra in Fig.~\ref{fig:cl_seds}, is the degree of irregularities in the two-hump SED shape. While in cluster~2 we observe a smooth spectrum, in cluster~3 the synchrotron hump becomes somewhat flattened due to enhanced emission at frequencies $\log_{10}\nu > 15$ in the rest frame, and in cluster~4 the average SED obtains a step-like shape.The statistical characteristics in Figs.~\ref{fig:distribs},~\ref{fig:luminosities} and Table~\ref{tab:cluster_stats} reflect the shape of the average SEDs. 
The source of the observed irregularities may be excessive emission from the central parts of AGNs.

Blazars from cluster~3  have sufficiently lower radio hardness parameters (Fig.~\ref{fig:distribs}).
Cluster~4 demonstrates noticeably higher median luminosities in the radio, X-ray, and gamma-ray ranges. Statistically, this cluster contains the most luminous objects at higher redshifts, if we do not consider few individual objects in cluster~2 with the greatest redshift and gamma-ray luminosity.

Distances are from 500 to 6500~Mpc ($z\sim 0.05$--$2.5$) for cluster~3 and from 2000 to 8000~Mpc ($z\sim 0.4$--$4$) for cluster~4.
For $z=4$ the frequency shifts in the rest frame are as follows: IR radiation goes into the optical range 3.35~{\textmu}m~$\Rightarrow$~6700~\AA; optical radiation goes into the UV range: 7520~\AA~$\Rightarrow$~1500~\AA; UV radiation becomes harder: 2316~\AA~$\Rightarrow$~460~\AA.

\section{Summary}

In this paper we discuss the applications of cluster analysis technique to the multi-wavelength properties of the blazars from the Roma-BZCAT catalog. 
We divided the blazars into five groups and compared them with the Roma-BZCAT classification, high synchrotron peaked blazars from the 3HSP catalogue and TeV blazars from TeVCat. We have found similar trends in blazar grouping, which confirms the effectiveness of the clustering technique. The obtained groups (clusters) 
are derived based on multiparametric distributions of blazar characteristics.

To perform the project, we collected data from the radio to gamma-ray ranges both from the Roma-BZCAT catalog itself and from various other point-source catalogs, mostly those containing sufficient amount of data for the sample. During clustering, the blazars were treated in the same manner regardless of the degree of our knowledge about them, e.g., we did not add any additional measurements from other catalogs for some well-known objects, thus preserving the homogeneous approach to the sample as a whole.

In general, clustering algorithms build an independent unsupervised 
classification that is based almost solely on the multiple properties of the objects under consideration.
In this sense, clustering is a more uniform and homogeneous approach to classify cosmic objects based on the experimental data avoiding subjective selection bias. The method, nevertheless, has its own hyperparameters (the parameters that is set by the researcher rather than learned by
the model from data): the feature space, which determines the characteristics relevant for the scientific scope of the problem; the algorithm implemented to find the clusters; the number of clusters, a trade-off between uniformity and individuality.

Here for the feature space we used the maximum available number of characteristics that could be related to the properties of the objects. Such an approach helps describe blazars in the most complete way possible. The process of feature selection is described in detail in Section~\ref{sec:feature_space}. In total, the model feature space comprises 14 continuously distributed characteristics. Although adding new parameters (for instance, because of the growing number of observations) will inevitably change the clustering results for particular objects, especially on the cluster boundaries, the overall patterns observed in the sample will preserve, unless the amount of new characteristics is comparable to the number of those in the original feature space. This allows us to discuss with a certain robustness the sample properties as a whole, which is confirmed by the comparison with the known blazar classifications. Cluster membership of any particular objects, though, must be considered with great caution and additional analysis. Moreover, as no localized groups are revealed in the feature space and due to the incompleteness of the Roma-BZCAT catalog, boundaries between the clusters are only conditional and might change for a more complete sample. We evaluated the influence of sample incompleteness and feature selection on the clustering results in Section~\ref{sec:robustness} and expect that our cluster labels will be preserved with a Rand index of 80\%\ after adding a sufficiently large amount of new information.

We have tested several clustering algorithms and finally settled on two of them: PCA+k-means and self-organizing maps (SOMs). The advantage of
the latter is that SOMs can restore possible nonlinearities in data distribution, while PCA dimensionality reduction is a more straightforward and interpretable method of linear algebra. By showing the 90\% similarity of their results, we demonstrate the absence of nonlinearities in our data as well as some robustness of feature space division into clusters: although the final stages use the k-means algorithm in both cases, they work in sufficiently different spaces, a 14D space of neuron weights in the case of SOMs and a 6D space of PCA components for the other method. As the methods in our case have shown similar results, we followed PCA+k-means for interpretation clarity.

The number of clusters is a hyperparamater that can be set rather loosely in the clustering problems. Generally, addition of a new cluster leads to division of an existing one into two subclasses. In the case of a continuous data distribution, the boundaries of clusters may also vary. 
We chose the number of clusters to be five based on the best match between data distributions within the clusters obtained for the subsample without missing values and for the whole sample where the missing values were imputed via probabilistic PCA. 

Speaking of the latter, we found it to be most effective for data imputation among other methods: imputation with medians or various machine learning regressions. Notice that these imputed values were only used to perform the clustering for the complete sample, we did not use them for the statistical analysis of the derived clusters.

The following notable characteristics of clusters have been derived.
\begin{description}
\item[Cluster 0] Consists of BL\,Lac-type blazars with low luminosities in all the ranges from radio to gamma rays except X-ray emission. Almost all known high synchrotron peaked blazars (HSPs) 
fall into the cluster. The synchrotron hump is not visible in the average SED as well as the second hump in the gamma-ray range, 
this effect is caused by the broad distribution of synchrotron peak frequencies in the cluster. The cluster is characterized by low radio luminosity, radio hardness parameters (Table~\ref{tab:cluster_stats}), and redshifts $z\lesssim0.9$.

\item[Cluster 1] Consists of BL\,Lac-type blazars with low luminosities in all the ranges from radio to gamma rays. The average SED is of the usual shape with two humps, but the gamma-ray hump is weaker than usual and comparable by flux densities with the synchrotron one. Low radio luminosities and radio hardness parameters. Redshifts $z\lesssim0.9$.

\item[Cluster 2] A mix of BL\,Lac-type objects (29\% including BL\,Lac candidates) and FSRQs (60\%). The rest of the objects (11\%) are of an uncertain type. High luminosities, strong radio hardness parameters. A clear smooth average SED with two humps. The entire range of redshifts.

We show that the BL\,Lacs from this cluster form a special subclass of high-luminosity BL\,Lacs compared to the low-luminosity population in clusters~0 and~1. 

\item[Cluster 3 and 4] FSRQs. High luminosities. The clusters are primarily distinguished between each other by the degree of irregularities in the SED shape that may be caused by the influence of the emission from the AGN central parts. Blazars from cluster~3 demonstrate statistically lower radio hardness parameters comparative to the FSRQs from clusters~2 and 4.
Cluster~4 has noticeably higher median luminosities in the radio, X-ray, and gamma-ray ranges, actually this cluster contains, on average, the most luminous objects at higher redshifts, although individual record holders as of the redshift and gamma-ray luminosity fall into cluster~2.
\end{description}

Our results are consistent with
the term ``blazar sequence'' originated in \cite{1998MNRAS.299..433F} to describe the properties of blazar SEDs. The most well-known feature of this phenomenological sequence is the negative correlation between the synchrotron peak frequencies and synchrotron peak luminosities of the blazar population, i.e., HSP blazars having the lowest luminosities and the highest synchrotron peak frequencies and LSP blazars with the opposite characteristics. But in a later study by \cite{2008A&A...488..867N}, that anti-correlation was not found in the intrinsic blazar properties after correcting the observed data for the Doppler beaming effects, and another study by \cite{2012MNRAS.420.2899G} also showed that the originally reported anti-correlation was due to a selection effect. Following works based on various multiband data just confirmed that the emissions in blazars is strongly beamed and affects the observational phenomenon known as ``blazar sequence'' (e.g., \citealt{2017ApJ...835L..38F,2023ApJ...949...52O,2024MNRAS.528.7529W}). In our study we did not apply the Doppler factor as one of the parameters because it is estimated for a limited number of blazars, e.g., for 979 Fermi blazars in one of the recent studies \citep{2024ApJS..271...20C}. Therefore, we cannot test the intrinsic nature of the blazar sequence.

\cite{2021MNRAS.505.4726K} studied a dichotomy in jets, dividing more than 2000 blazars into two samples: one with inefficient accretion weak/type~I jets and the second with efficient strong/type~II jets. The first group contained blazars with synchrotron peak frequencies above $10^{15}$~Hz (HSPs, nearly all BL\,Lacs), and the second one comprised mostly FSRQs and some LSP BL\,Lacs. This quite accurately coincides with our findings both by synchrotron peak frequency values and blazar type distribution, i.e., clusters~0 and 1 contain blazars with type~I jets, and clusters~3 and 4 are the blazars with type~II jets.

We believe that these groups of Roma-BZCAT blazars, derived as a result of multiparametric analysis, can be used as additional information for further research, for example in the search for correlation with neutrino events or other statistical investigations.

The dataset with various characteristics of the blazars and cluster labels is available in the VizieR database (CDS).

\section*{Acknowledgments}
We thank an anonymous referee, whose valuable comments helped improve the paper. This study was funded by the Ministry of Science and Higher Education of the Russian Federation under contract 075-15-2022-1227.
The research has made use of the Roma-BZCAT blazar catalog and references therein, Astrophysical CATalogs support System (CATS), NASA/IPAC Infrared Science Archive, NASA/IPAC Extragalactic Database (NED), Barbara A. Mikulski Archive for Space Telescopes (MAST), SED Builder, Sloan Digital Sky Survey (SDSS), and Two Micron All Sky Survey (2MASS).
The publication has made use of data products from the Wide-field Infrared Survey Explorer (WISE), Panoramic Survey Telescope and Rapid Response System (Pan-STARRS), Galaxy Evolution Explorer (GALEX), RATAN-600, FIRST, Planck, ROSAT, Swift-XRT, Fermi, and a number of other telescopes the observed data of which have been compiled in Roma-BZCAT, CATS, and SED Builder.
We used a set of Python libraries that are mentioned in the text.

\bibliography{ms2024-0055}{}
\bibliographystyle{raa}

\end{document}